\documentclass[%
 reprint,
 amsmath,amssymb,
 aps,
floatfix,
]{revtex4-1}

\usepackage{graphicx}
\usepackage{dcolumn}
\usepackage{bm}
\usepackage{hyperref}

\begin{document}


\title{Early Gumbel forecasts of the evolution of the COVID-19 outbreaks\\ and quantitative assessment of the effectiveness of countering measures.}

\author{Emanuele Daddi$^1$}
 \email{edaddi@cea.fr}
\author{Mauro Giavalisco$^2$}%
\email{mauro@umass.edu}
\affiliation{%
$1$) CEA, IRFU, DAp, AIM, Universit\'e Paris-Saclay, Universit\'e Paris Diderot,
Sorbonne Paris Cit\'e, CNRS, F-91191 Gif-sur-Yvette, France
}%
\affiliation{$2$) Department of Astronomy, University of Massachusetts Amherst, 710 N. Pleasant St., Amherst, MA 01003, USA}


\date{\today}

\begin{abstract}
We discovered that the time evolution of the inverse fractional daily growth of new infections, $N/\Delta\, N$, in the current outbreak of COVID-19 is 
accurately described by a universal function, namely the two-parameter Gumbel cumulative function, in all countries that we have investigated. While the two 
Gumbel parameters, as determined bit fits to the data, vary from country to country (and even within different regions of the same country), reflecting 
the diversity and efficacy of the adopted containment measures, the functional form of the evolution of $\Delta\, N/N$ appears to be universal. The 
result of the fit in a given region or country appears to be stable against variations of the selected time interval. This makes it possible to robustly 
estimate the two parameters from the data even over relatively small time periods. In turn, this allows one to predict
with large advance and well-controlled confidence levels, the time of the peak in the daily new infections, its magnitude and duration (hence the total 
infections),  as well as the time when the daily new infections decrease to a pre-set value (e.g. less than about 2 new infections per day per million 
people), which can be very useful  for planning the reopening of economic and social activities. We use this formalism to predict and compare 
these key features of the evolution of the COVID-19 disease in a number of countries and provide a quantitative assessment of the degree of 
success in in their efforts to countain the outbreak.
\end{abstract}

\maketitle


\begin{figure*}[ht]
\begin{centering}
\includegraphics[width=8cm]{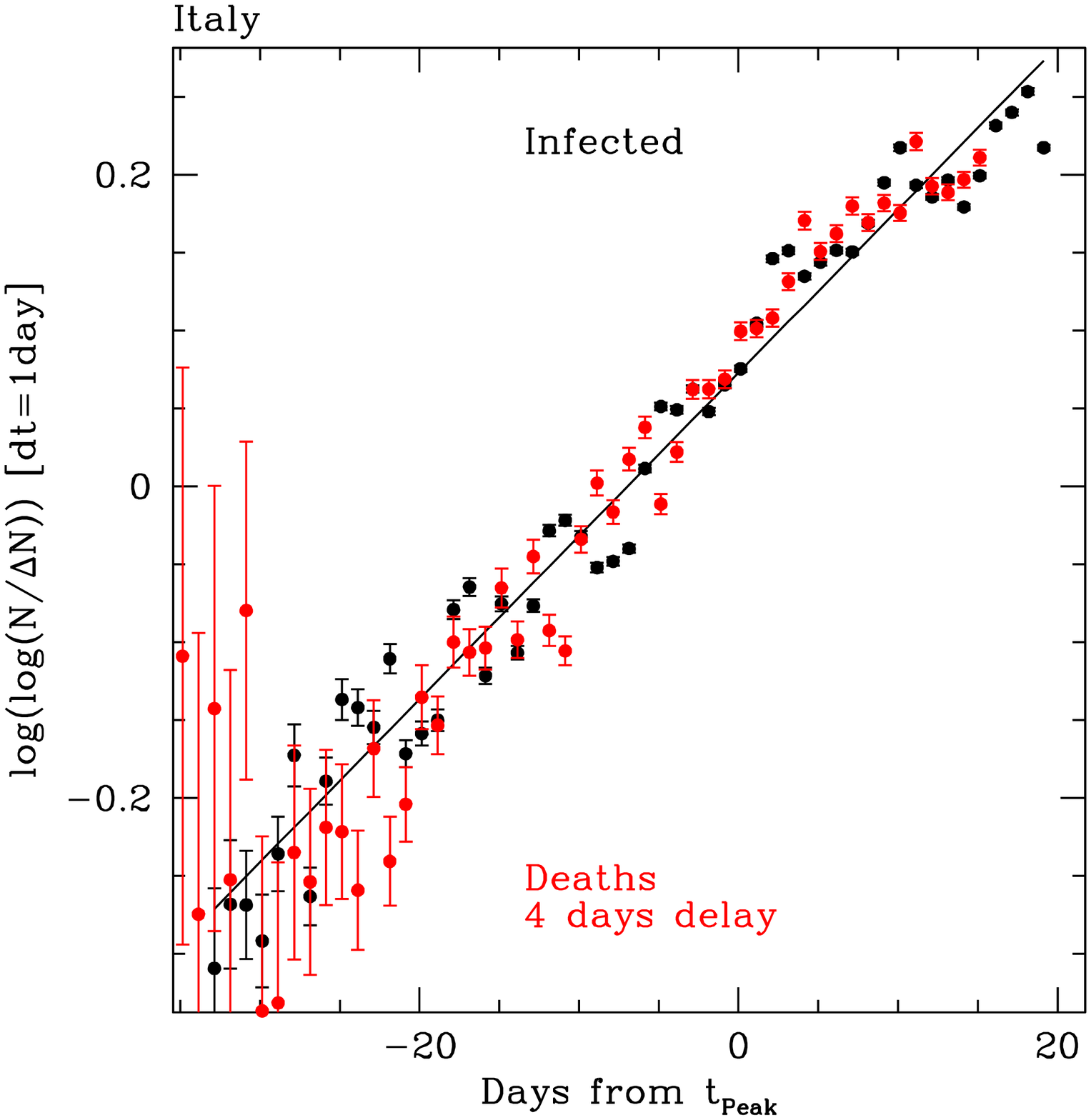}
\includegraphics[width=8cm]{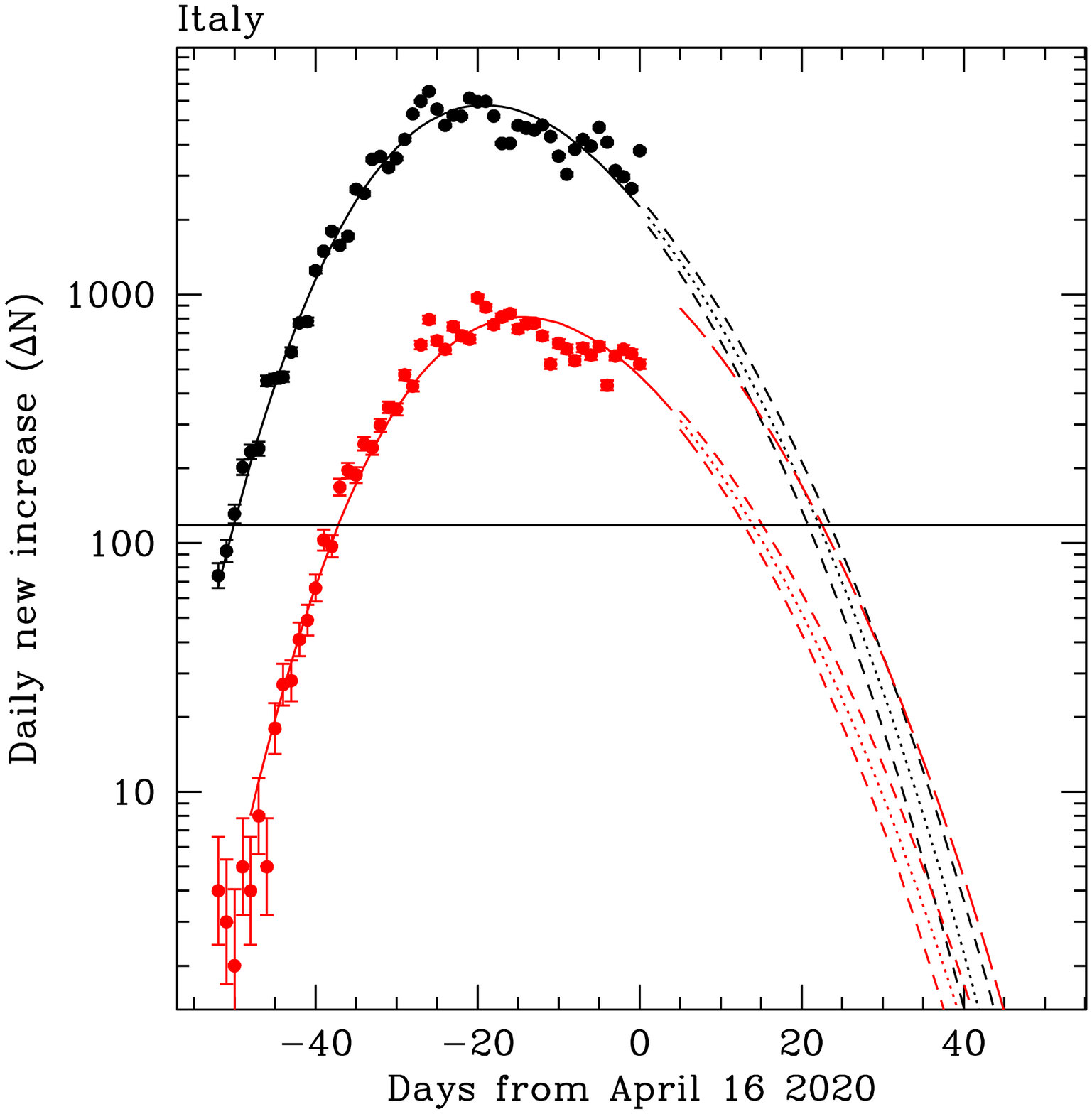}
\caption{\label{fig:ita1} {\bf Left:} The double log of the ratio of total infected to newly infected is plotted as a function of time (black points). 
The line is the best fitted Gumbel function, as defined in Section~\ref{sec:Gumbel}. In the case of a pure exponential ($N\propto e^{t/\tau}$), this ratio is equivalent to the characteristic evolutionary timescale $\tau$ (if the growth from one day to the next were to be considered exponential). The increase in the ratio, as shown in this plot, is thus a direct visualisation of the deceleration in the spread of the disease. Red points shows the same quantities for deaths, delayed by 4 days: it can be seen that deaths follow the same evolutionary rate with typical delays that were of 6-7 days at early times and get to 3-4 days more recently, due to the increasing number of recovered patients (deaths are a more accurate delayed function of $N_{\rm active}$, but exploring this is beyond the scope of this work).
{\bf Right:} We show daily values of $\Delta N$ for infected (black) and deaths (red, delayed as in the left panel).  The solid curves correspond to the Gumbel best fitting function from the left panel (delayed for deaths and scaled down by the integrated CFR); dotted line shows the extrapolation of the same function; dashed lines show the 1$\sigma$ range around the forecasted trend. The red long-dashed line shows the extrapolation of the sum of deaths and recovered: when this long-dashed line rises above the dotted black line, $\Delta N_{\rm active}$ becomes negative and $N_{\rm active}$ thus starts decreasing, hence corresponding to the peak of $N_{\rm active}$. Throughout this work we refer instead to the peak of $\Delta N$, as shown in this figure. The horizontal line shows the level of 2/day/1M new infected, that we set as the floor to a regime where the outbreak can be controlled as seen in South Korea.
}
\end{centering}
\end{figure*}

\begin{figure*}[ht]
\begin{centering}
\includegraphics[width=4.5cm]{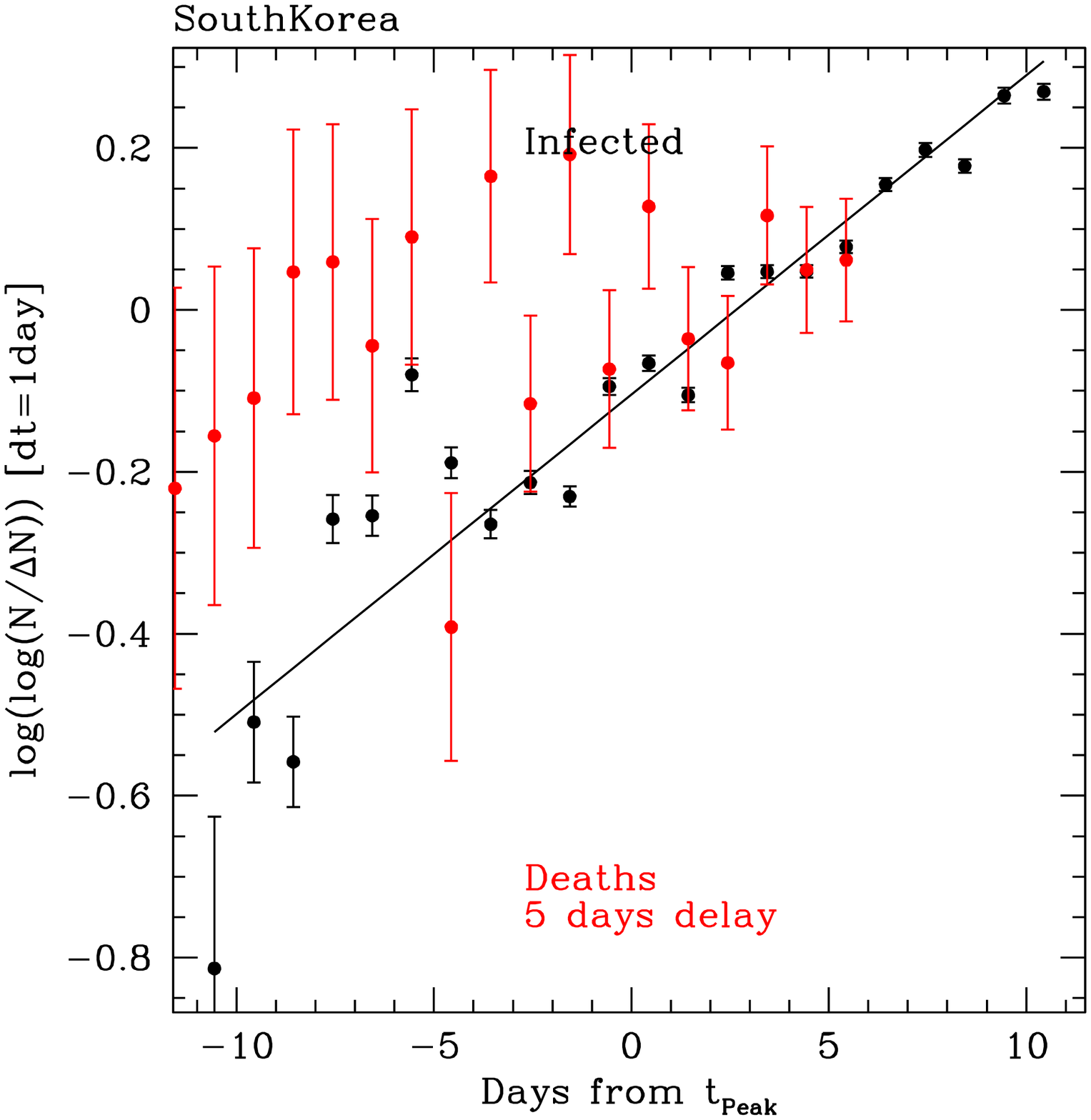}
\includegraphics[width=4.5cm]{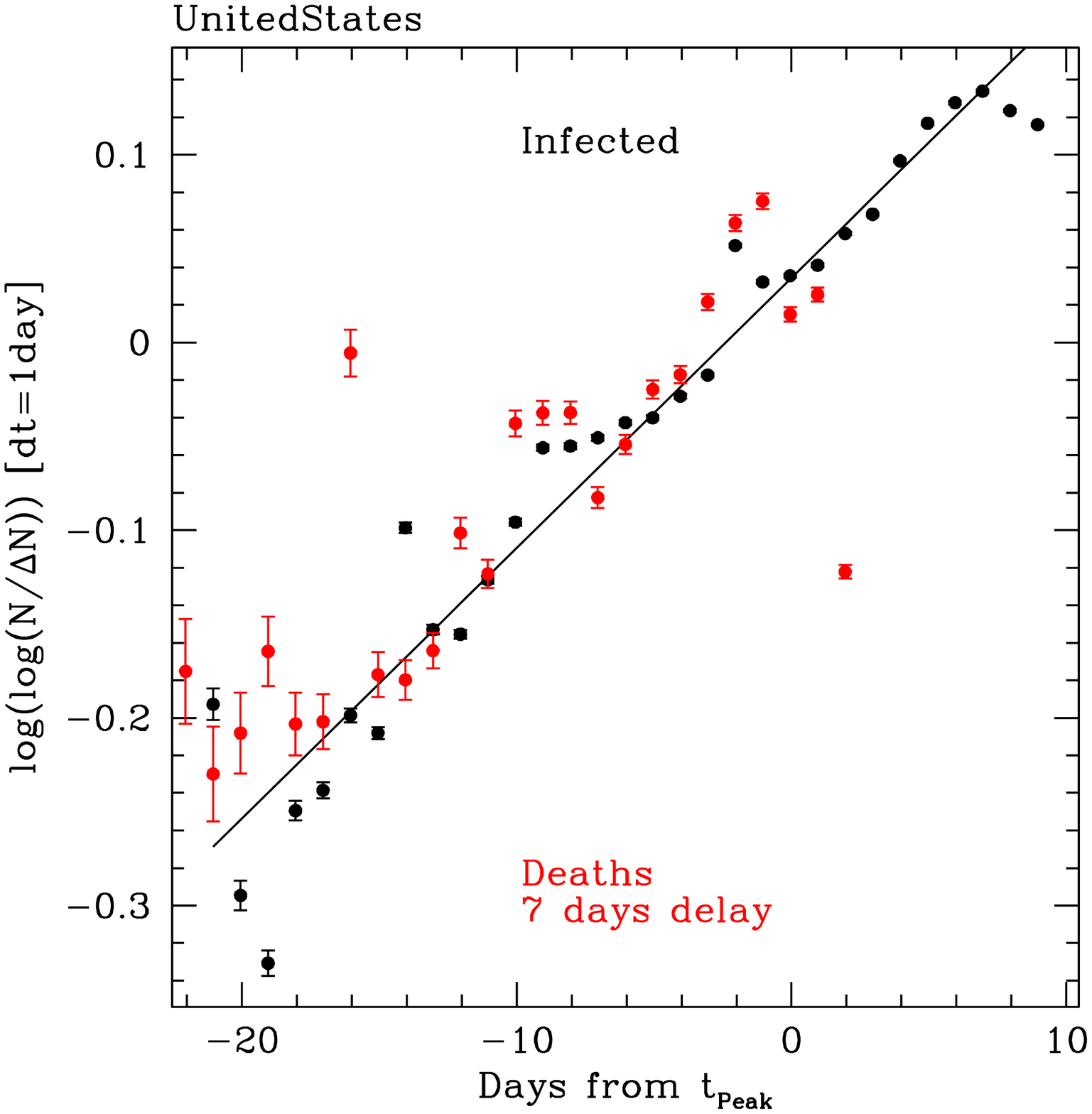}
\includegraphics[width=4.5cm]{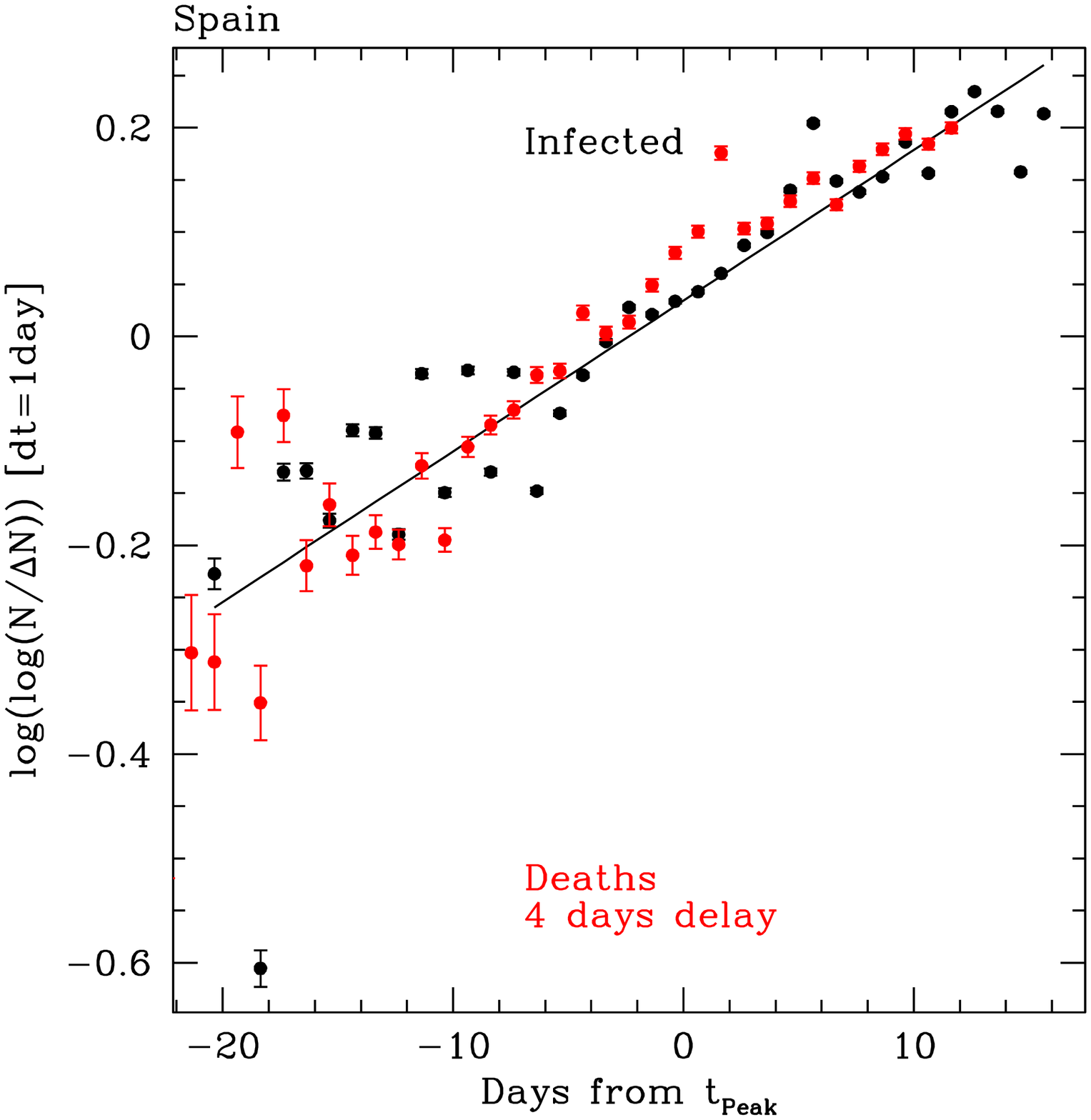}
\includegraphics[width=4.5cm]{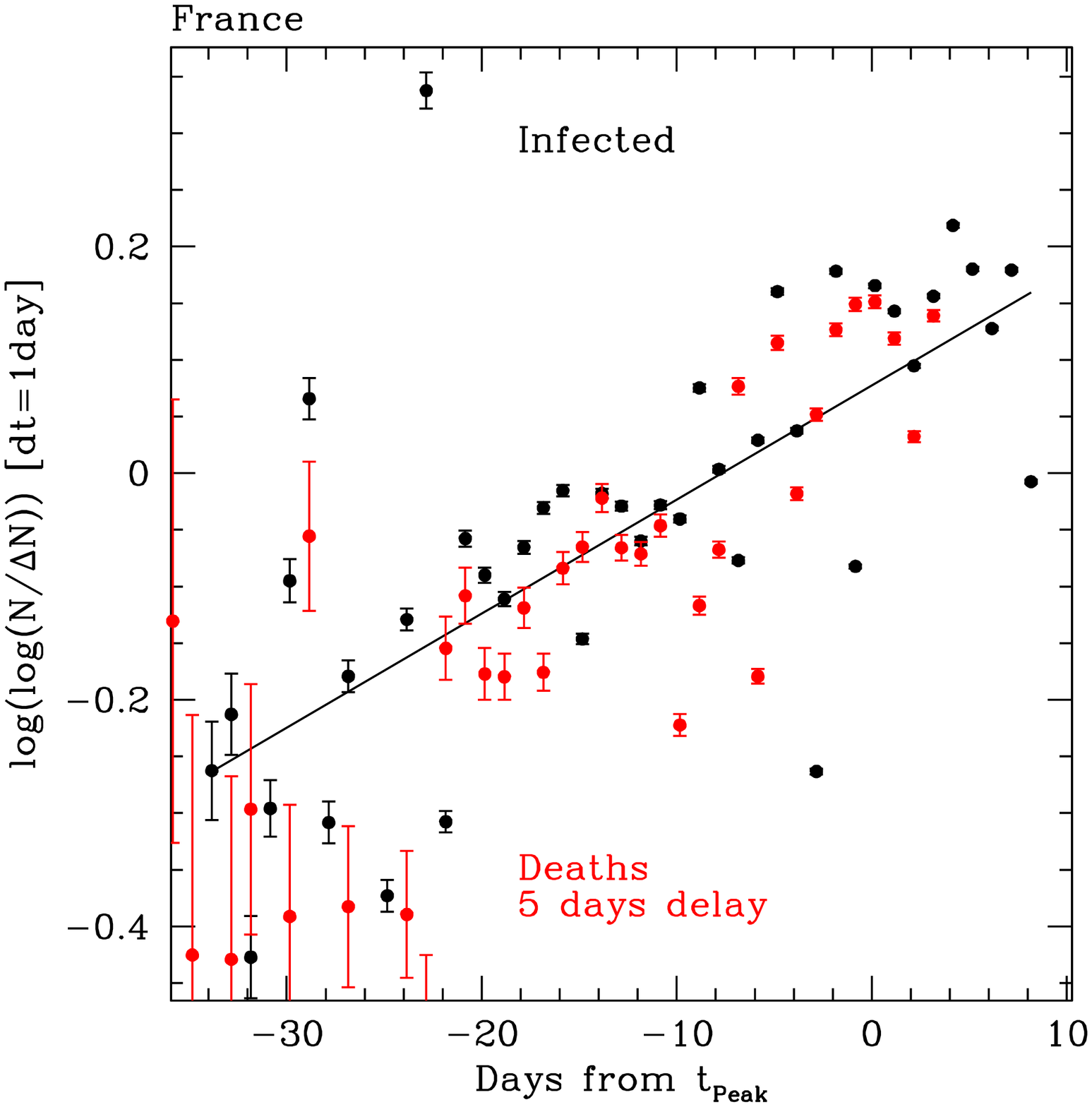}
\\
\includegraphics[width=4.5cm]{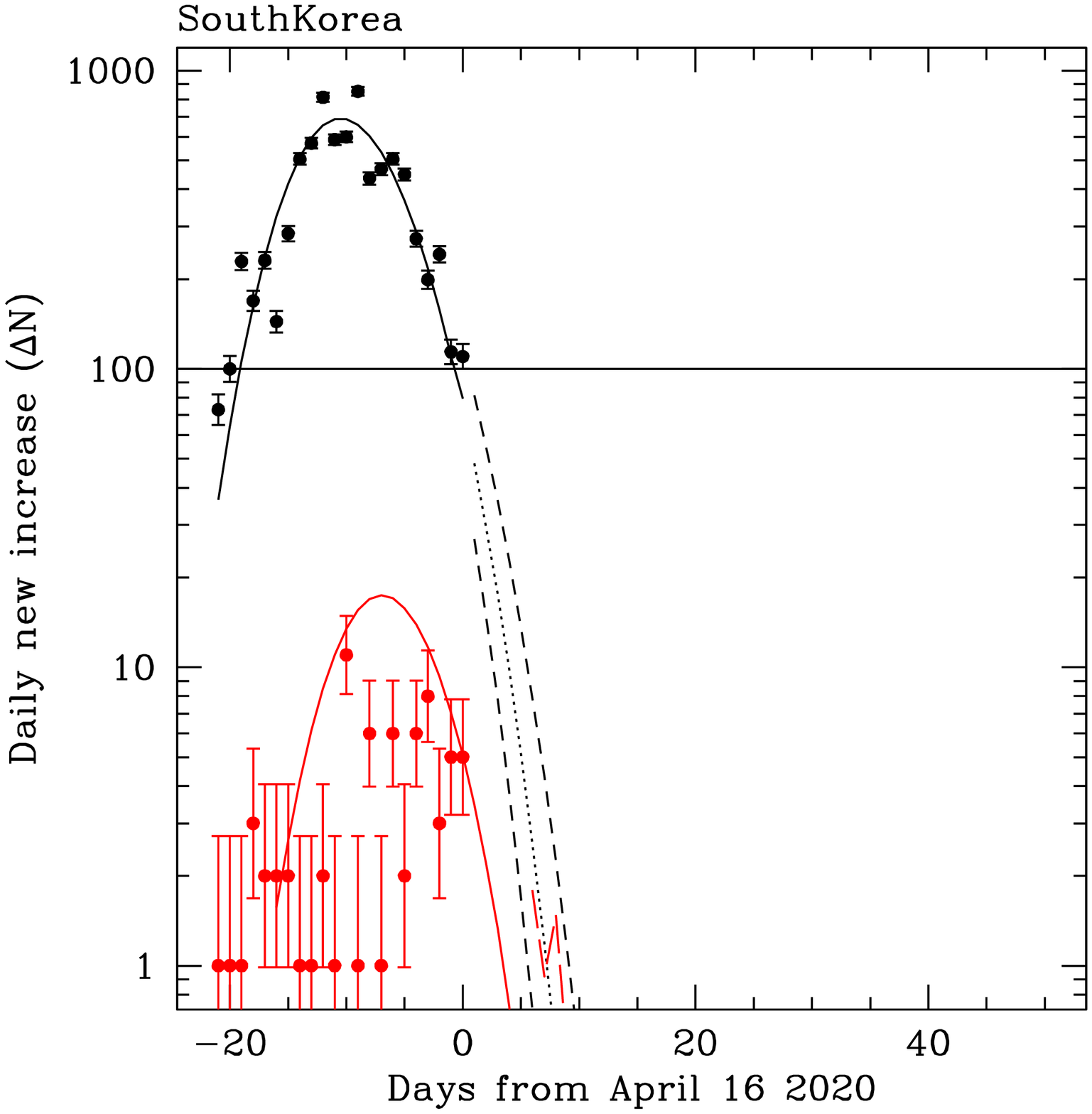}
\includegraphics[width=4.5cm]{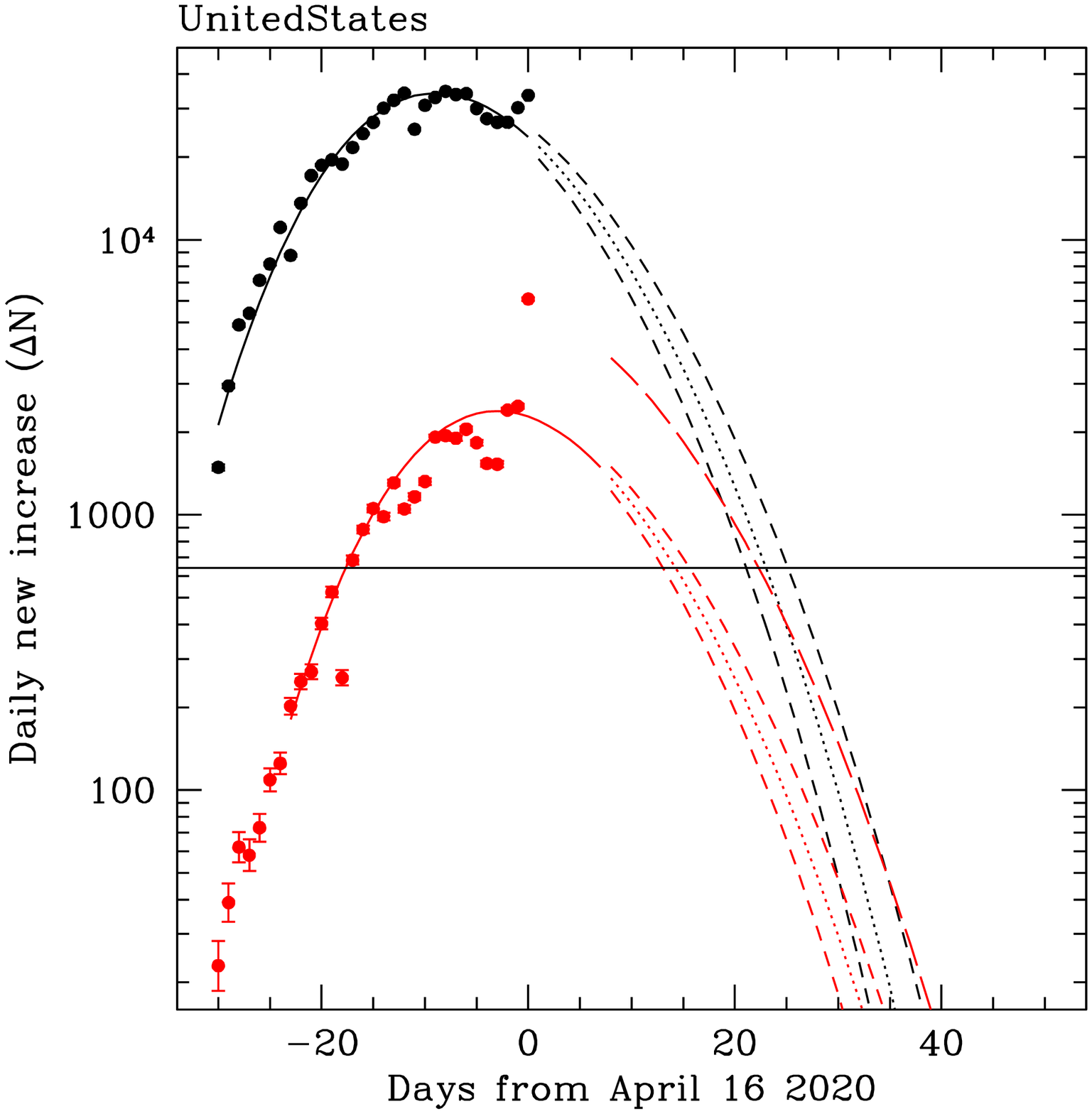}
\includegraphics[width=4.5cm]{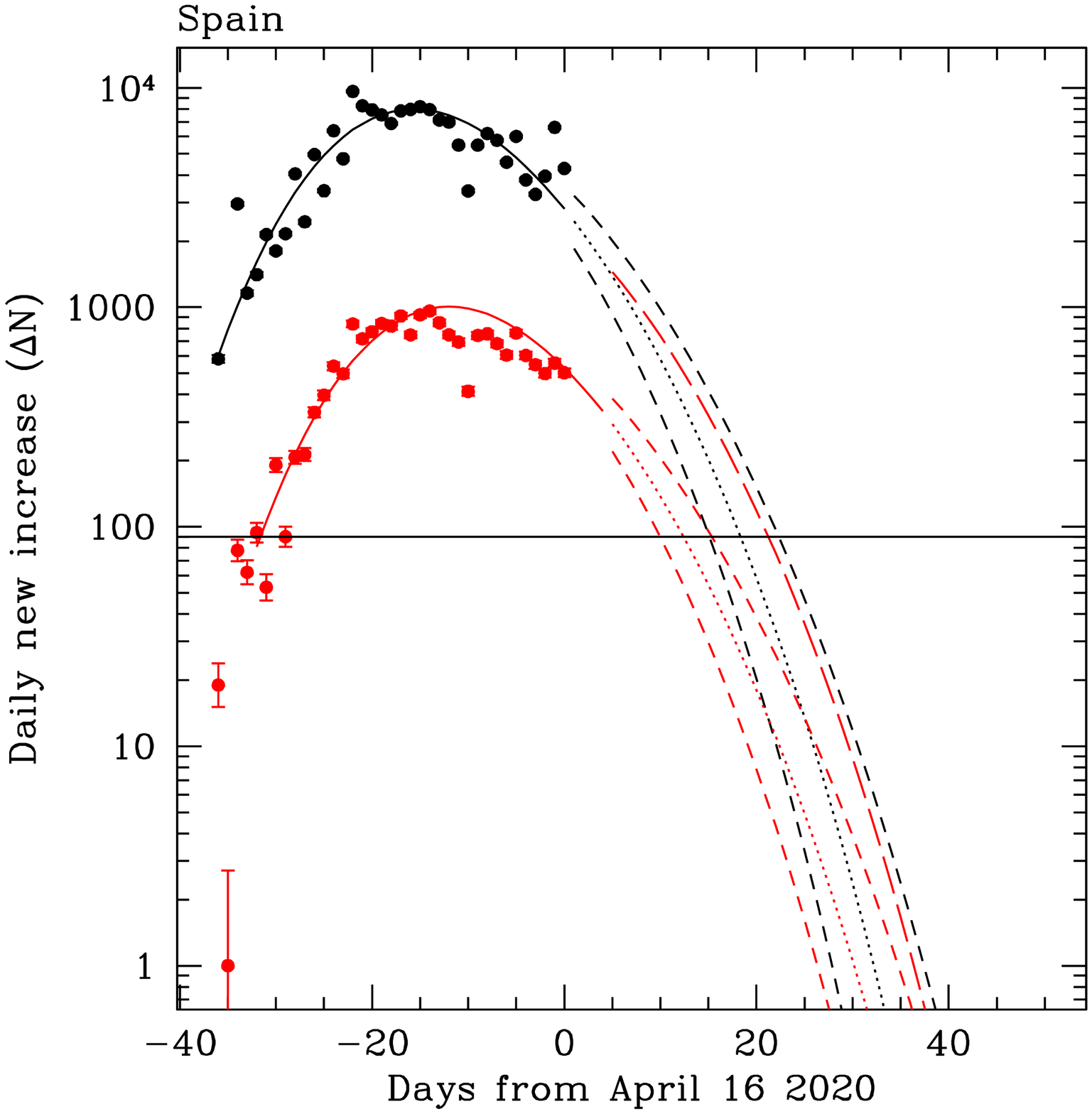}
\includegraphics[width=4.5cm]{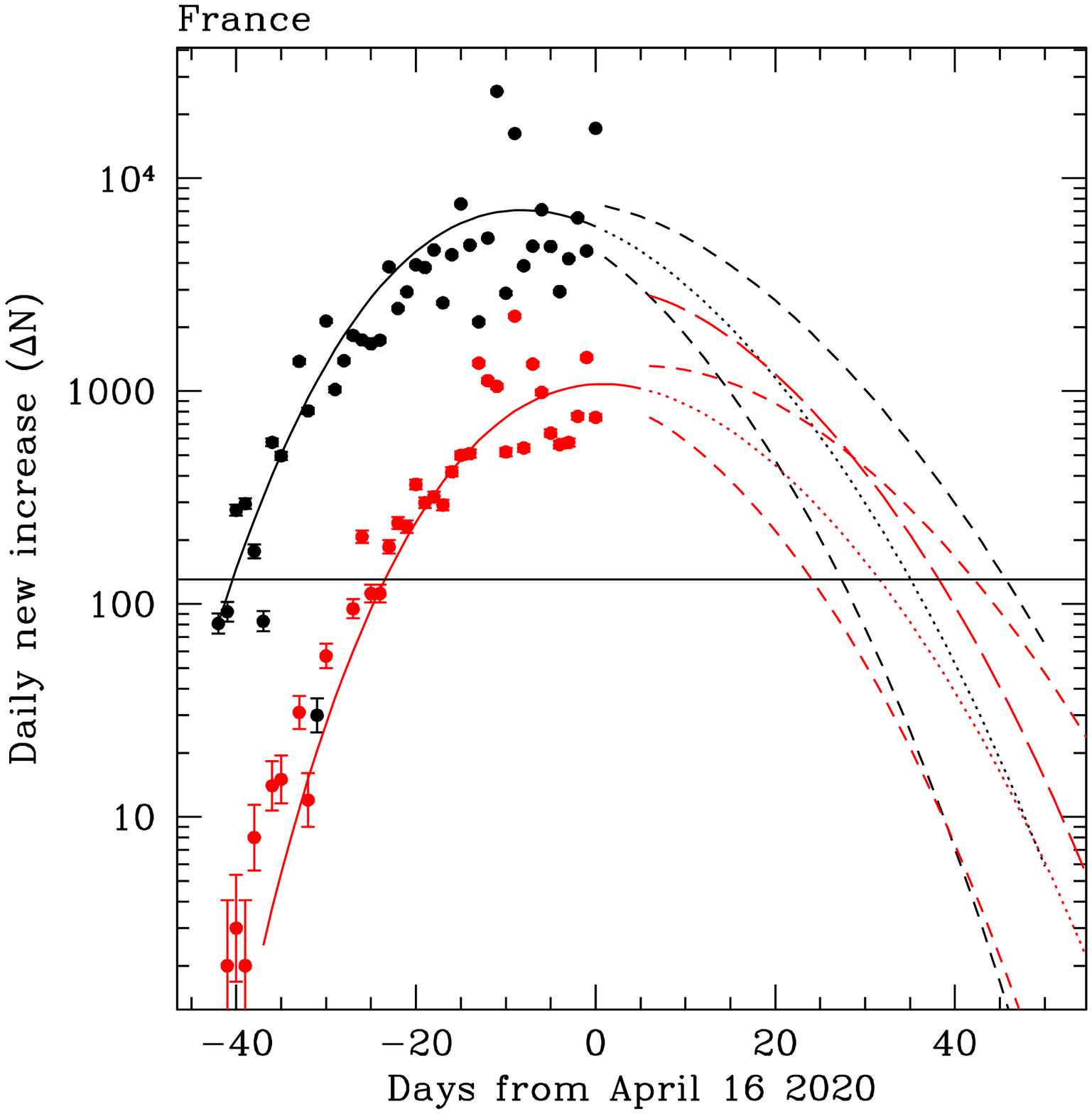}
\caption{\label{fig:more}  The same of Fig.~\ref{fig:ita1} but for South Korea, the US, Spain and France, as labeled.
}
\end{centering}
\end{figure*}

\begin{figure*}[ht]
\begin{centering}
\includegraphics[width=4.5cm]{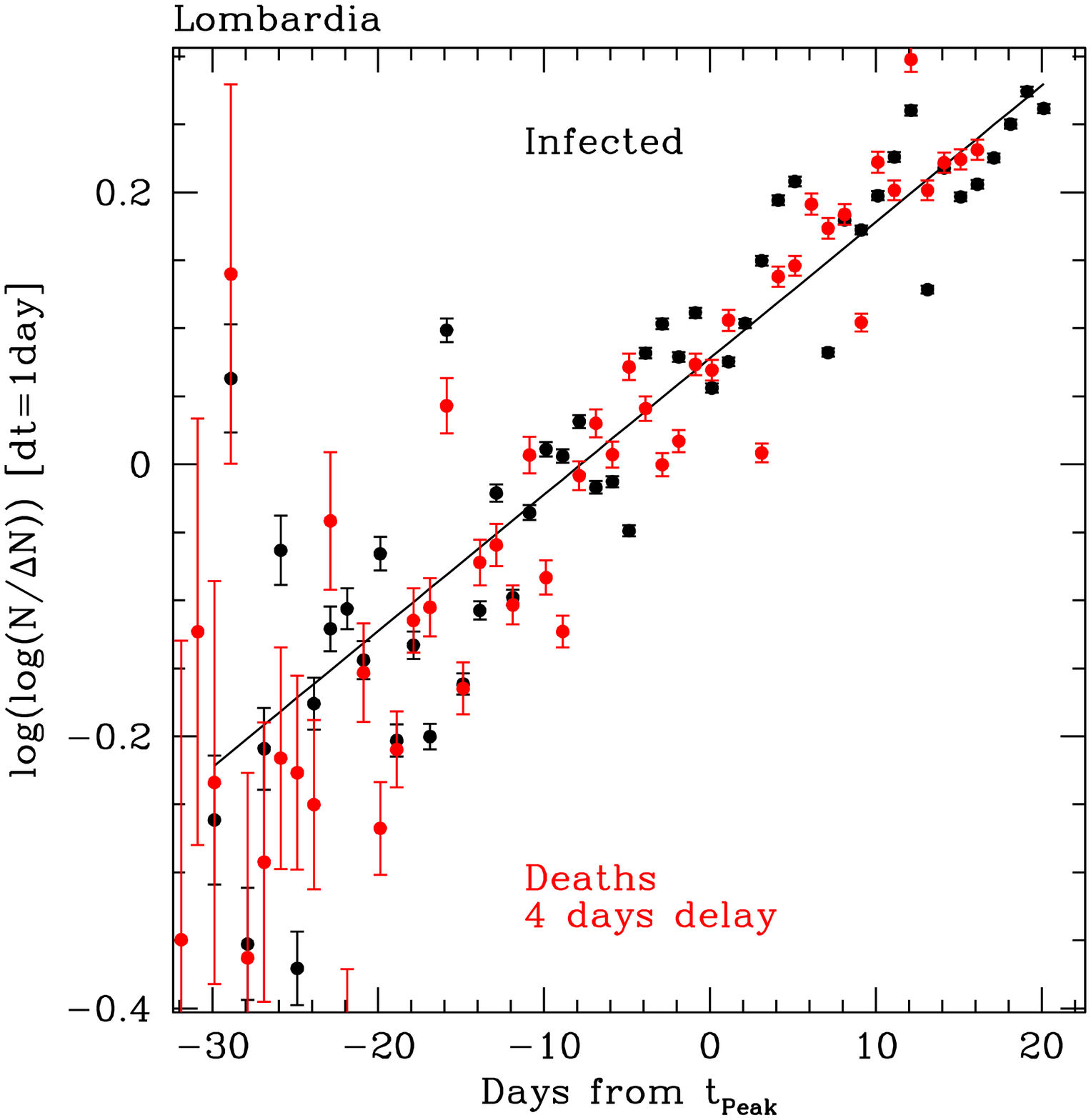}
\includegraphics[width=4.5cm]{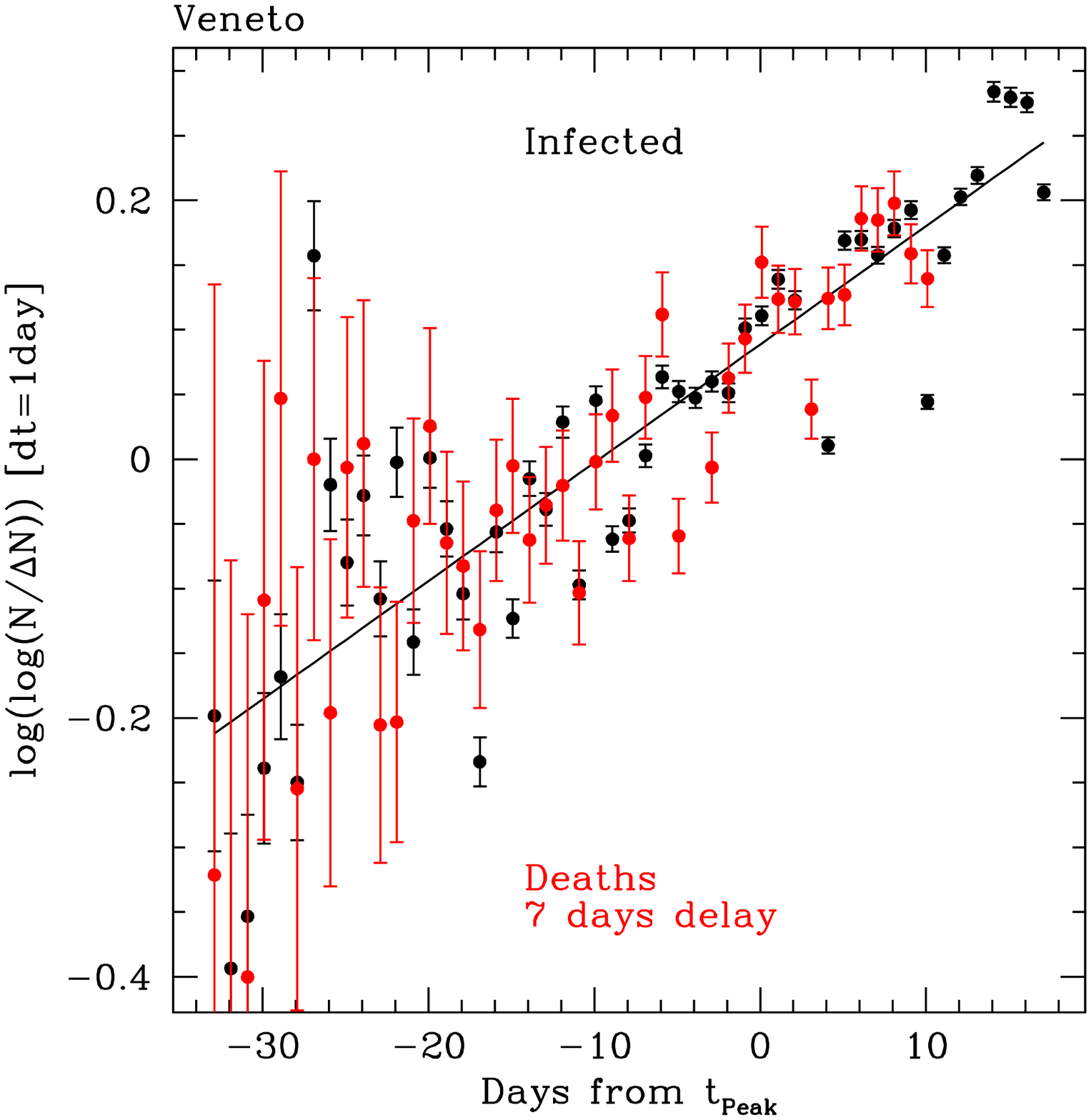}
\includegraphics[width=4.5cm]{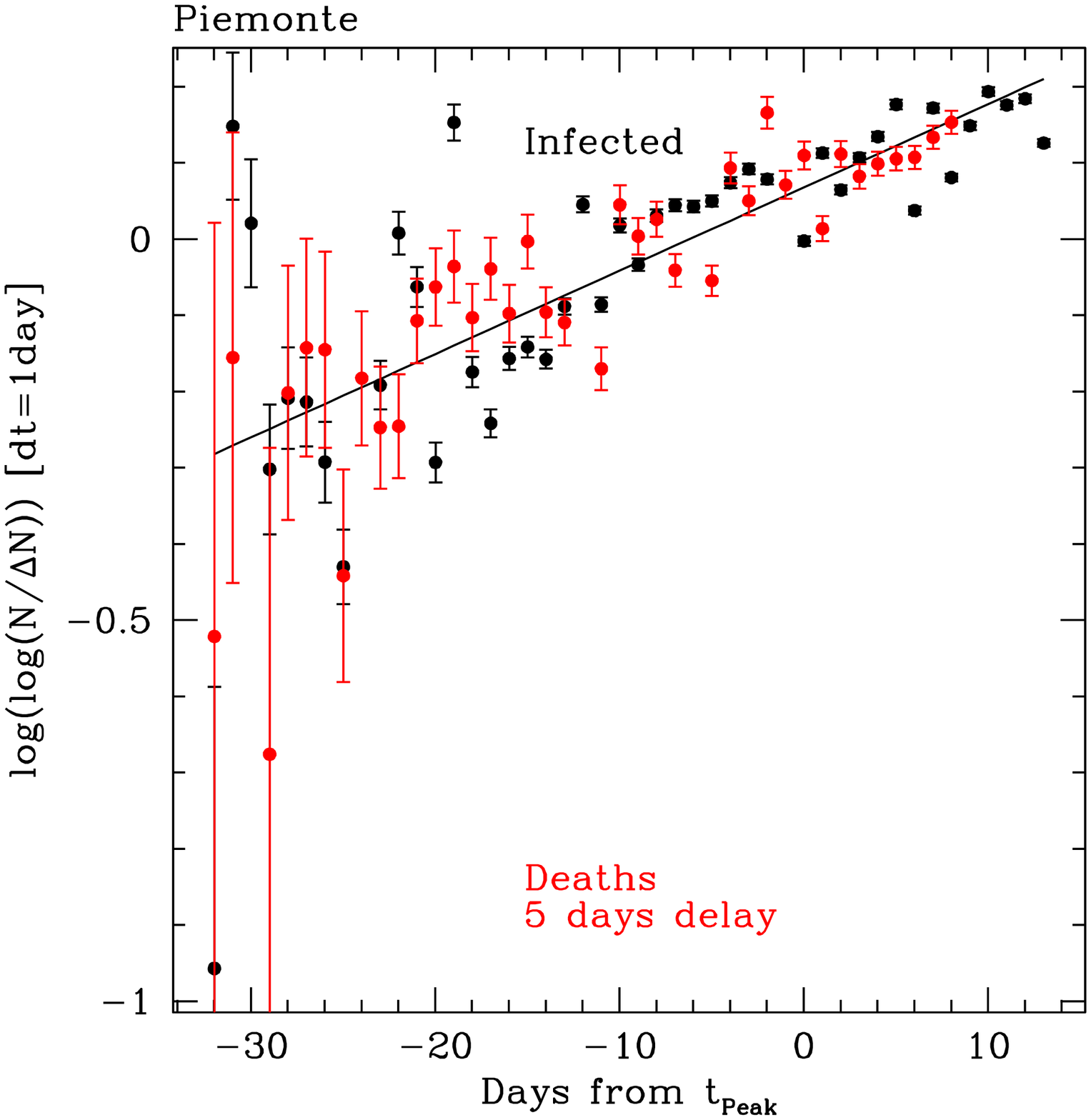}
\includegraphics[width=4.5cm]{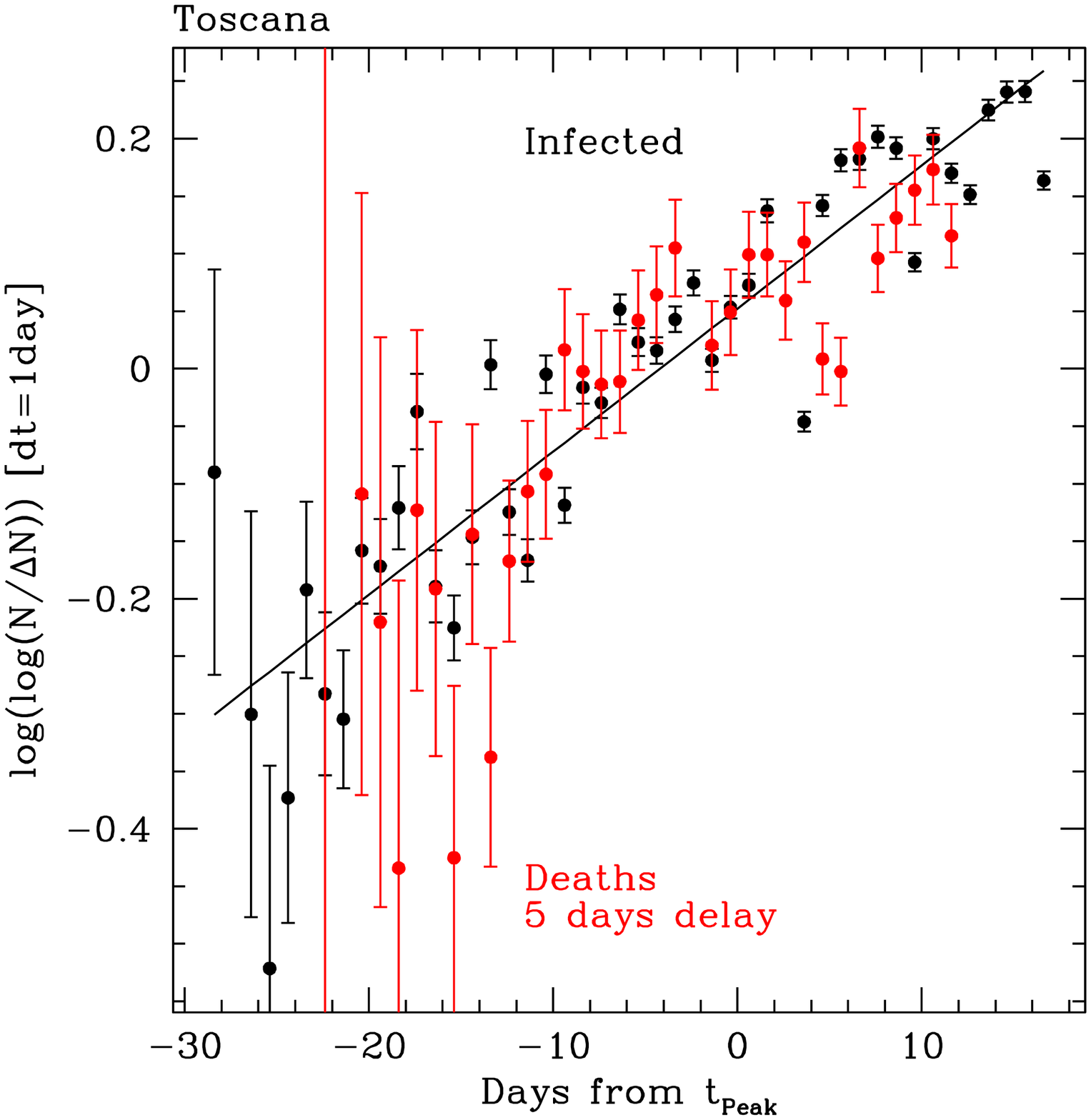}
\\
\includegraphics[width=4.5cm]{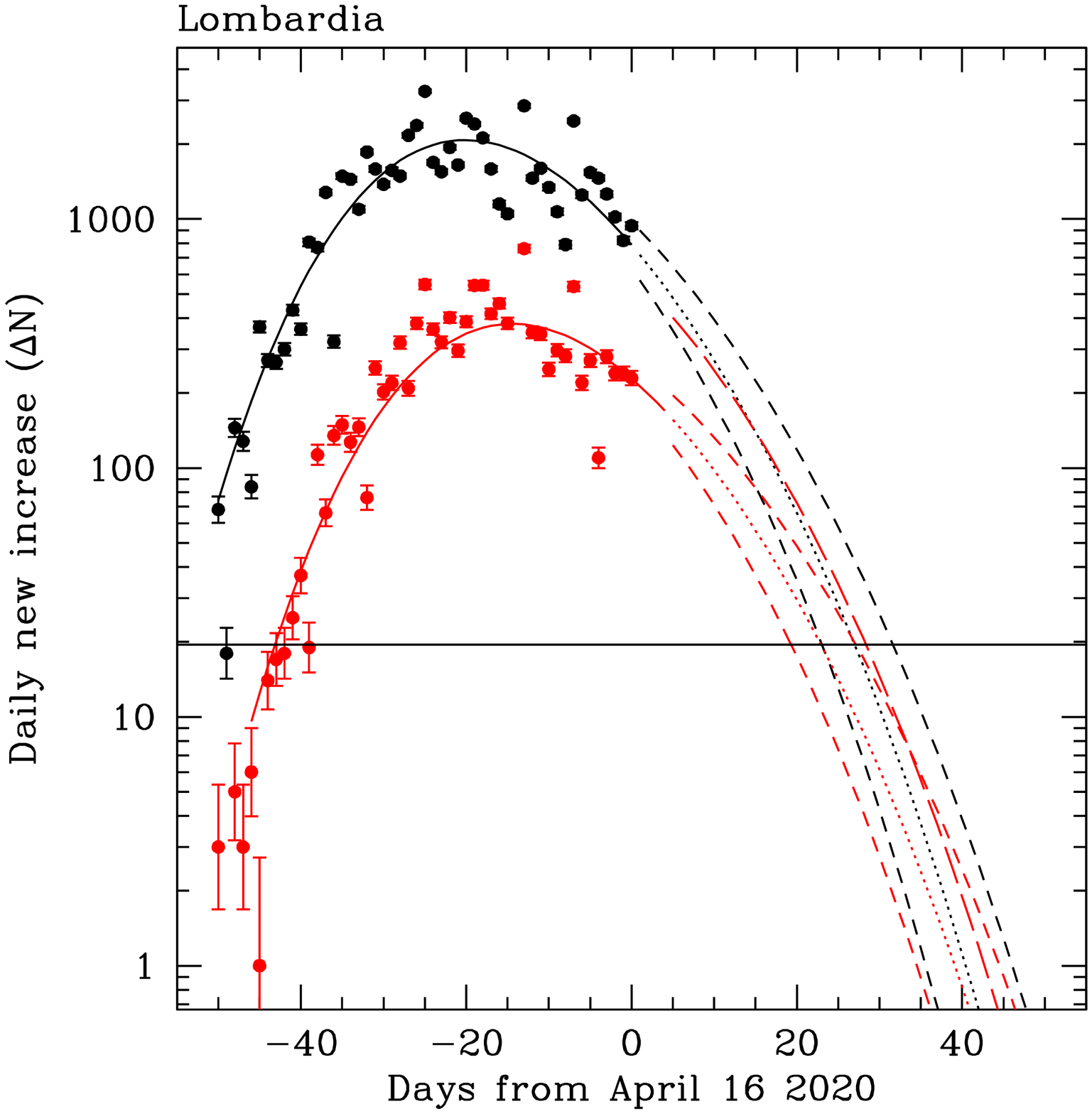}
\includegraphics[width=4.5cm]{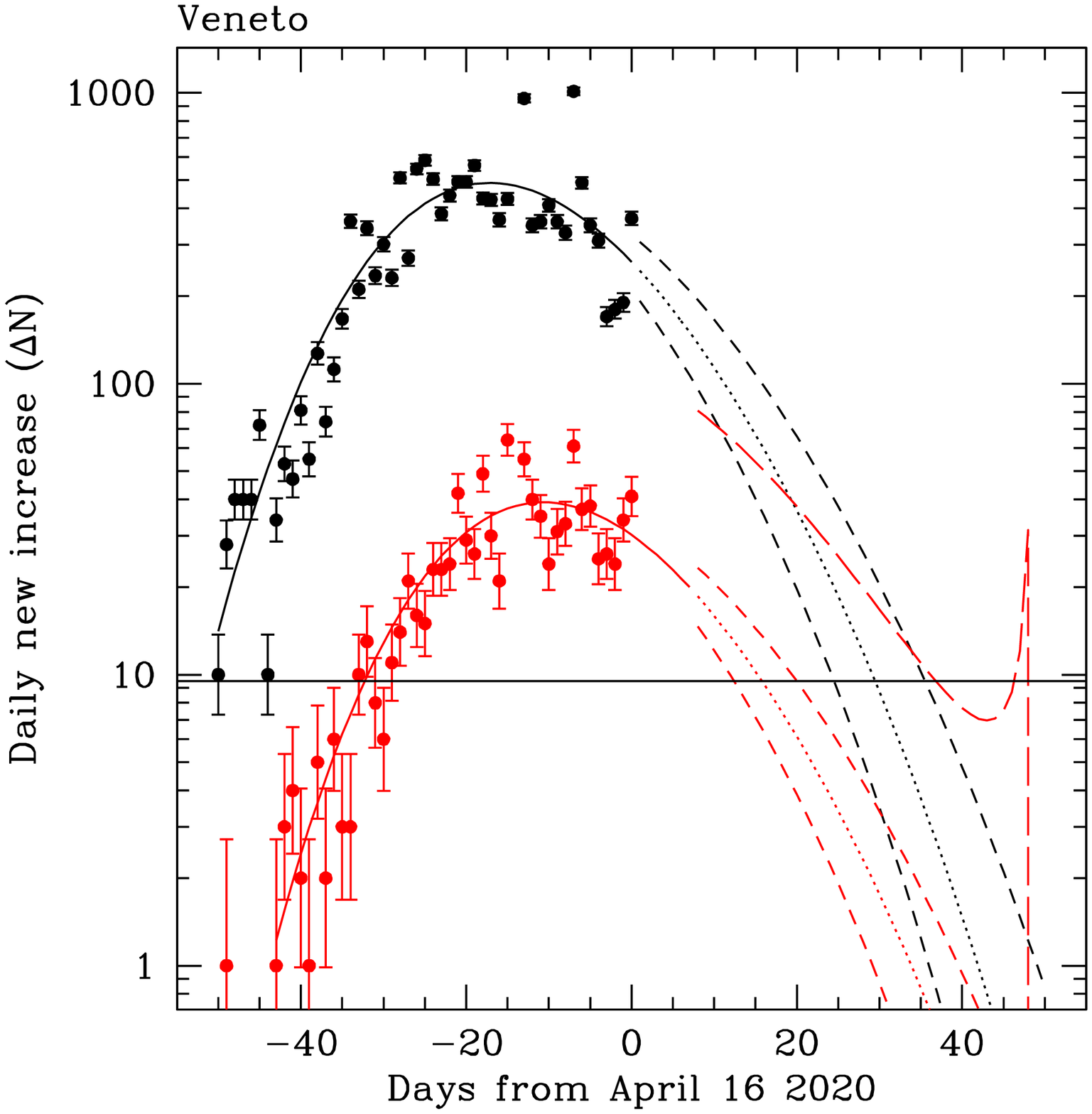}
\includegraphics[width=4.5cm]{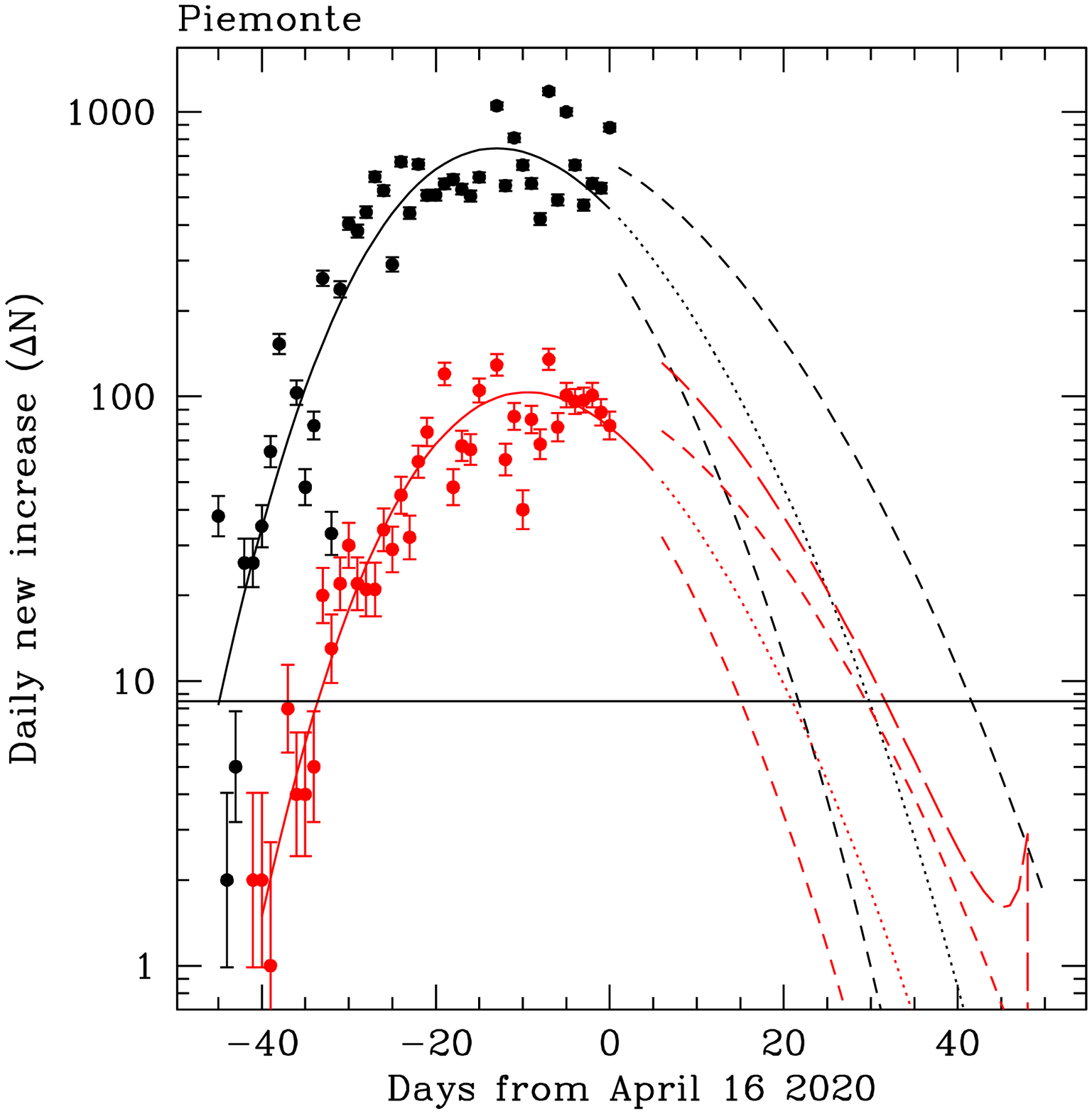}
\includegraphics[width=4.5cm]{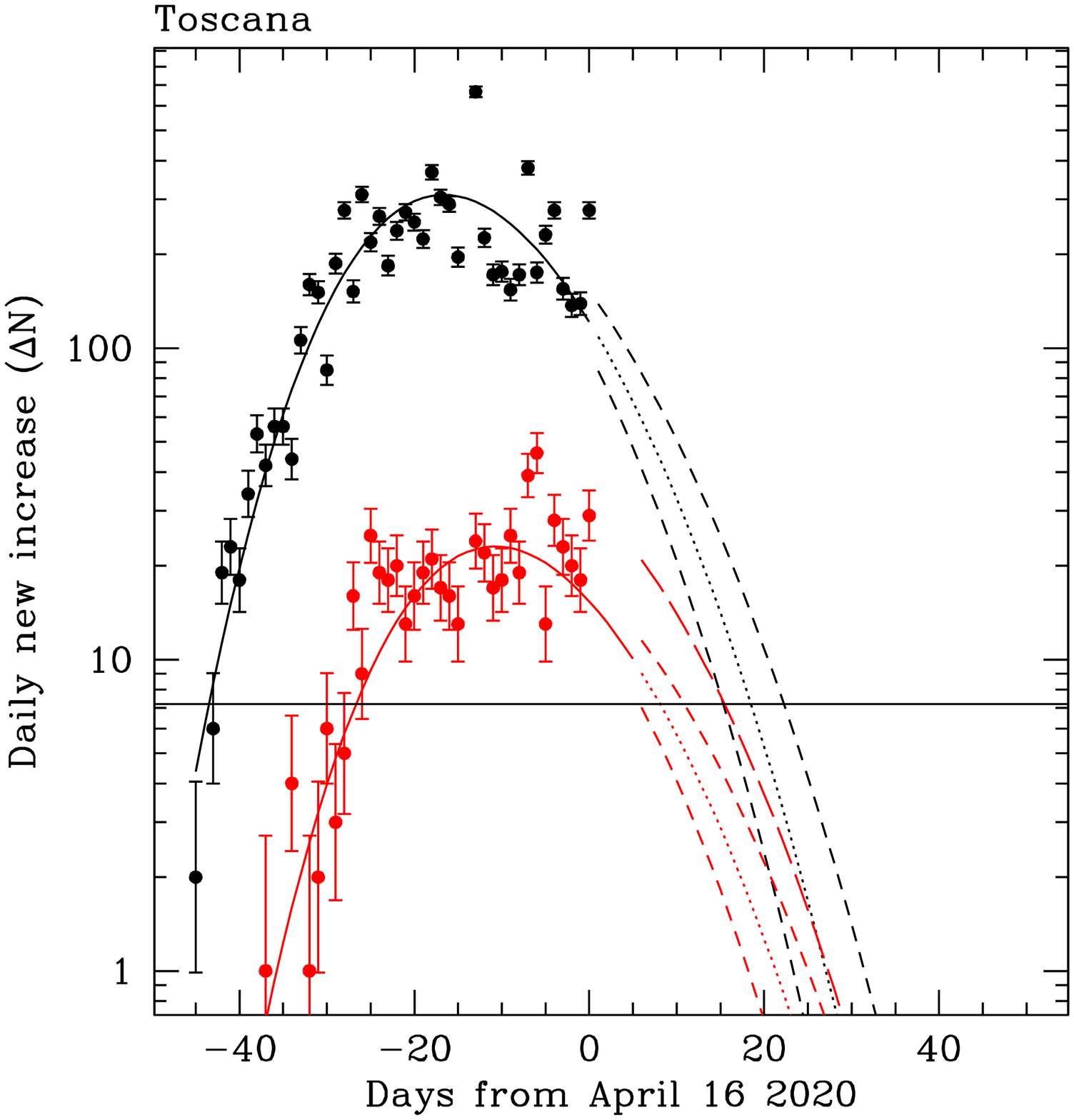}
\caption{\label{fig:regions}  The same of Fig.~\ref{fig:ita1} but for various regions of Italy, as labeled.
}
\end{centering}
\end{figure*}

\section{Introduction}

In absence of containment measures, the early phase of an pandemic outbreak, such as the current COVID-19 (ref. [1--9]) is characterised by an 
exponential growth with time of the number of infected people, $N(t)$  ([28-33]. In this phase, the rate of new infections as a function of time, 
$\dot N(t)=dN/dt$, such as the number of daily new infected 
$\Delta\, N(t)$,  is exponential as well. If left unimpeded, the epidemic naturally evolves and slows down, with $N(t)$ departing from the initial 
exponential growth and $\Delta\, N$ reaching a peak value before starting to decrease. This happens when a sufficiently large fraction of the initial 
population has become infected that the number of new infections become progressively more inefficient and is matched and eventually surpassed 
by the rate of deaths and recovered individuals. Such an outcome, however, is not acceptable in the case the current COVID-19 outbreak, since it 
would result in a very large loss of life. Containment measures are necessary to limit the deaths and avoid saturating health care capabilities, their merit 
and efficiencies have been discussed in a number of recent studies as soon as the COVID-19 epidemics started spreading (cfr. Refs.~[1--9]). 
Given 
the relatively long incubation period of the disease, from 2 up to 28 days, and the fact that a large fraction of the infected population appears to be 
asymptomatic or hypo-symptomatic, however, how can a government monitor the effectiveness of the adopted containment measures and adjust 
them so that the number of new infected reaches a peak as quickly as possible? How can one predict when $\Delta\, N$ will peak  and when its 
decrease will begin and reach a low, pre-set value of production of new infected to be manageable? A substantial effort is already ongoing to attempt 
modelling
the number of various observables, including newly infected individuals, deaths and recoveries, and using a variety of methods from machine learning to the use of standard SIR approaches (cfr. Refs.~[10--18]). 
The goal of this paper is to introduce and 
describe a simple {\it phenomenological} model that can be effectively used to predict the time of arrival of the peak, and its width, solely based on the 
observed time evolution of the number of known newly infected individuals $\Delta\, N(t)$. The model is interesting in practice because it is based 
on the data themselves, without the need of any further assumption. It is also  conceptually attractive because, as we shall see, the data show that, 
in presence of containment measures, the functional dependence with time of the key observable $\Delta\, N/N$, the fractional number of daily new 
infected, appears to be universal and characterized by only two parameters, $\mu$ and $a$, as we will describe later. In other words, the presence 
of containment measures changes the growth of the epidemic from exponential into another functional form, which is universal, independently of the 
types and magnitude of the measures adopted by various countries to oppose the spread of COVID-19. These measures determine the parameters 
of the functional form of $\Delta\, N/N$ but not its general expression. In fact, even the exponential function can be considered as a special case of our general function, 
in the limit in which one of its parameters (the evolutionary timescale) goes to infinite. 

This work is organized as follow: in Section~II we introduce the mathematical framework and fitting methods. In Section~III we  describe the meaning of the Gumbel parameters and some of its properties. We further verify the forecasting capability of the method in Section~IV. In Section~V we evaluate quantitatively the performances in countering the outbreak by various countries. We discuss our results in Section~VI and conclude in VII. 

\section{Mathematical framework}
\label{sec:Gumbel}

Let's call $N(t)$ the total number of known infected people at a given time $t$. We define:

\begin{equation}
    dN = r N dt
\end{equation}

\noindent
where $r$, which we call "the reproductive rate per unit time", at any given time, is related to the $R_0$  parameter (generally introduced in so-called 
SIR models) by the relation $r=(R_0-1)\, \gamma$ (where $\gamma$ in SIR models is the inverse of the average duration of the infectious period). In an unchecked 
epidemic outbreak, $r$ and $R_0$ are constant and the outbreak grows exponentially until a significant fraction of the population has been infected. If containment measures are present, however, 
 $r$ is not a constant even in the early phases but is a a function of time as a result of the measures put in place to slow and halt the outbreak. 

Empirical analysis of the best studied case of the time evolution of $r$ in the COVID-19 outbreak, for example in Italy, Spain, South Korea 
and the US, shows that a rather accurate description of the observed data is provided by the  Gumbel 
(1935; [19]) cumulative distribution function: 

\begin{equation}
  dN/dt/N = r = e^{-e^{(t-\mu)/a}}
  \label{eq:gumbel}
\end{equation}

\noindent
as Figure 1 and 2 illustrate (note that this actually is a reflected Gumbel, given that $a$ is positive). This function, which is specified by only 2 
parameters that are determined by fits to the data, provides an excellent representation of the observations in all cases that we have considered, 
namely countries where the type, implementation and timing of the preventive measures are rather different. As the case of Italy shows, the Gumbel 
function provides an excellent representation of the data for the country as a whole, and also for the data of individual regions of Italy as well (cfr. 
Figure 1 and Figure 3). Since the fractional rate of new infected individuals is modeled as a double exponential, the implication is that $r$ has its 
own "reproductive rate per unit time" which is itself a function of time and decays exponentially, following the equation:

\begin{equation}
    dr = -r e^{(t-\mu)/a} dt.
\end{equation}

Fig.~\ref{fig:ita1}-left shows the Gumbel fit for Italy over the last 45 days (here and in the following we are using $dt=1$ day). We have collected 
data from {\it Worldometer} [21] and official Github repositories [22; 23], and included them in our analysis as they reached a daily production of new infected close to the 
1--2/day/M in each country.

To determine the parameters of the Gumbel function we perform a least-square fit of a linear function to the logarithm of the logarithm of $N/\Delta N$ with 
no weights, thus equally counting each data point, regardless of the Poisson errors, since it can be readily seen from the figure that the scatter of the 
points around the best-fit function is fairly constant in double-log space as a function of time (except perhaps very early phases that appear to be noisier). 
This suggests that the uncertainty of each measurement is likely dominated by fluctuations in the reporting and sampling systems rather then by 
counting (Poisson) errors, and therefore it must be related to the effectiveness and of the testing and reporting capabilities of each countries, as 
discussed later. We  tried different approaches such as fitting the cumulative Gumbel function to the $N$ time series, and the differential Gumbel 
function to the $\Delta N$ time series, but none provided as good a fit as with the one we found with our approach. This conclusion, we believe, is already 
apparent in the $N/\Delta N$ time series plots in Figures~\ref{fig:ita1}--\ref{fig:regions}, where there is no hint that the functional form of the double log of 
these series ever significantly departs from a linear relation. We notice that a recent study (Bianconi et al 2020; [20]) discuss modeling of the doubling time of newly infected patients as a function of time, which is equivalent to the quantity $r$ in our study (it's inverse). They argue that two asymptotic behaviours are present at early and late phase of the outbreak with different slopes in log space. When using a Gumbel function the need for ad-hoc change of slopes is removed, and the data evolution remain linear in double-log space.

\begin{figure*}[ht]
\begin{centering}
\includegraphics[width=4.5cm]{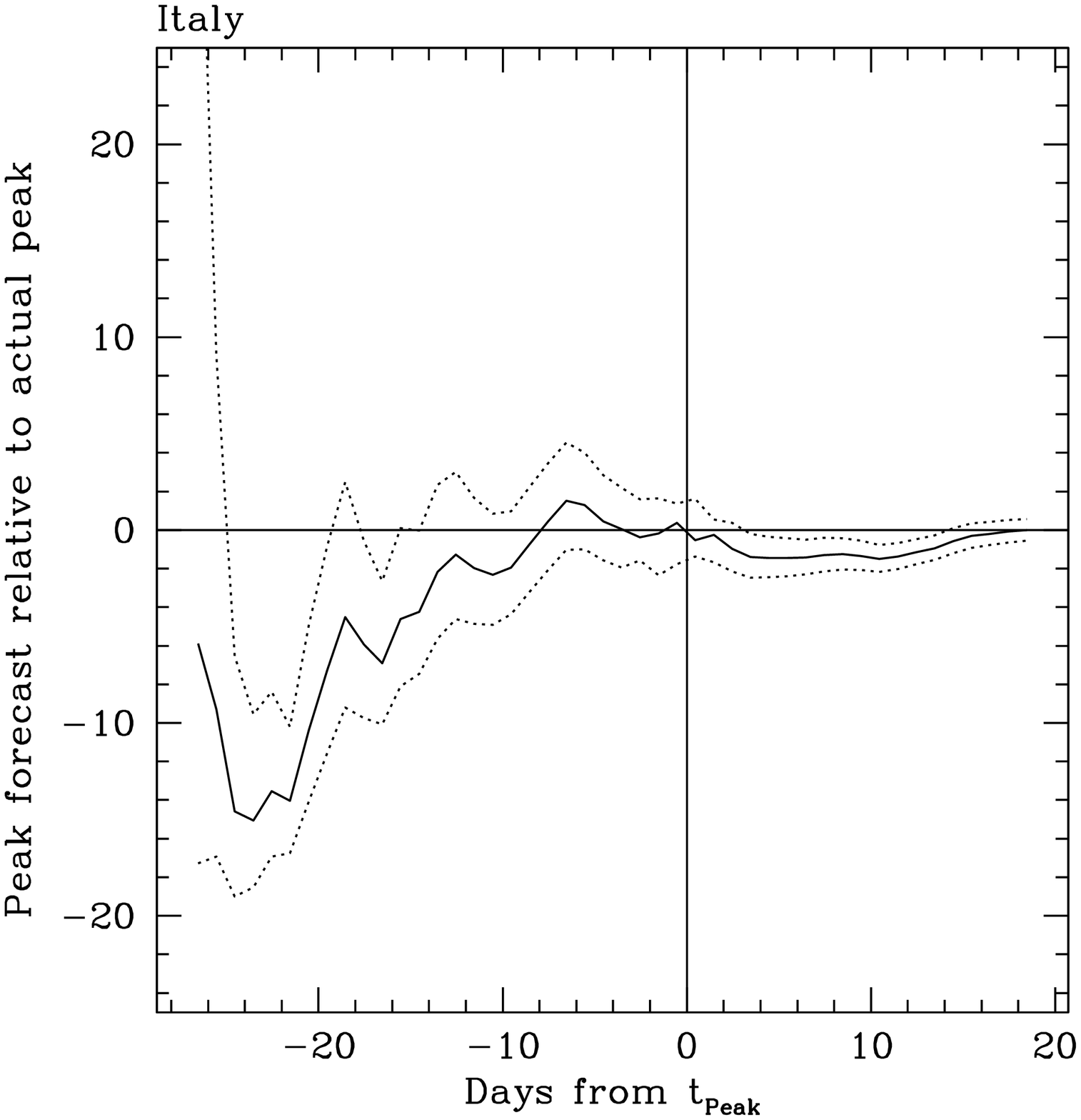}
\includegraphics[width=4.5cm]{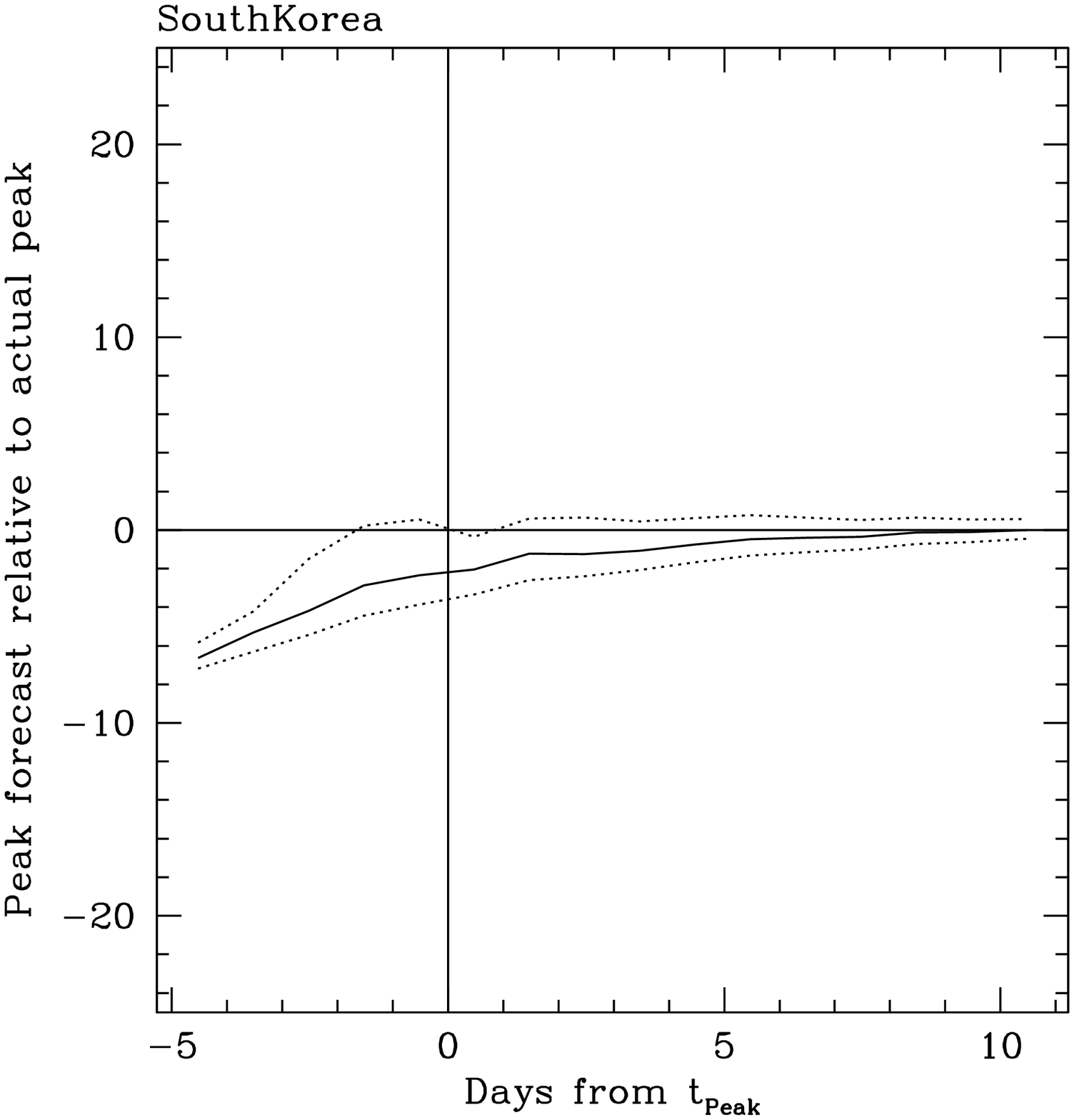}
\includegraphics[width=4.5cm]{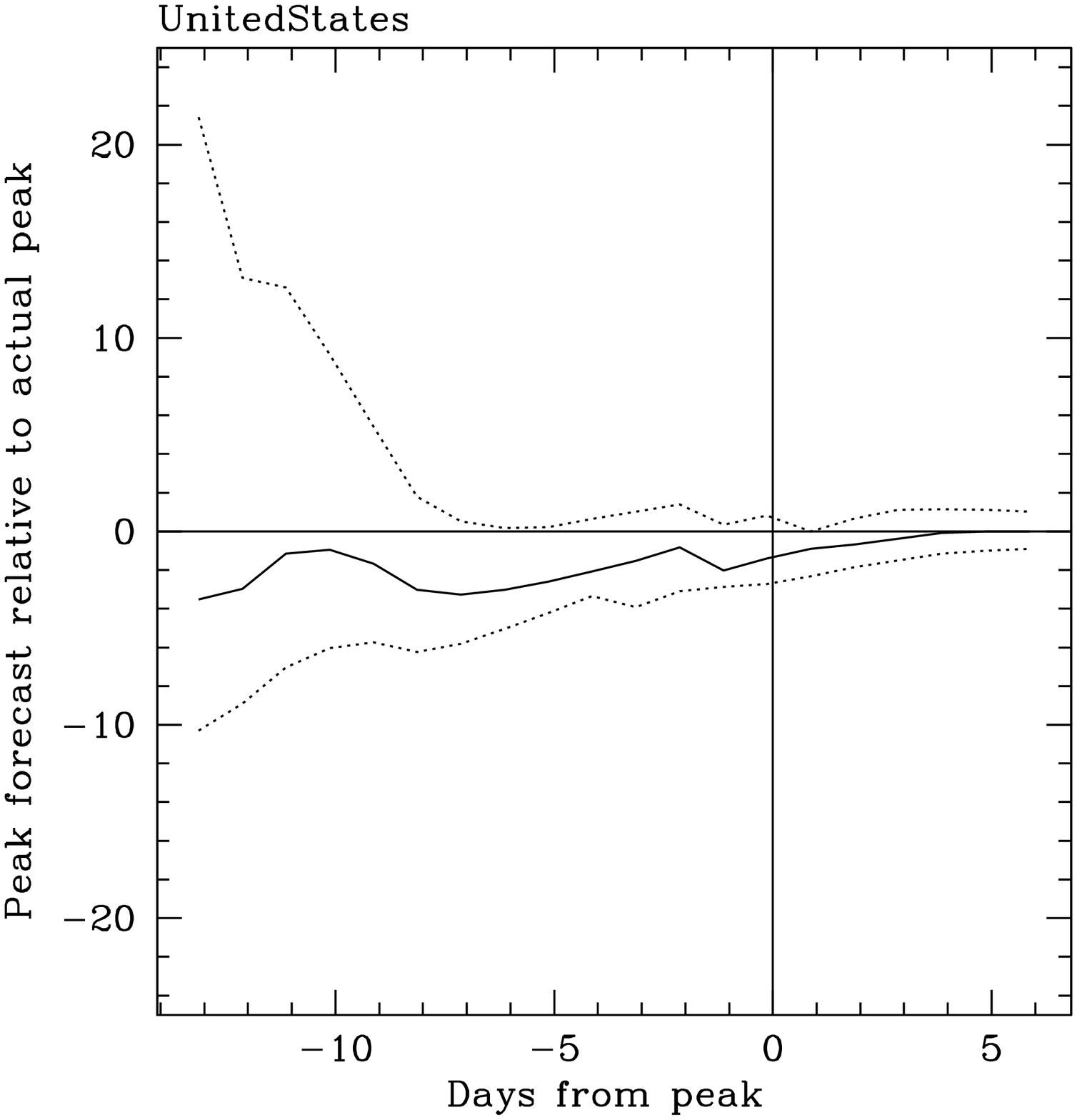}
\includegraphics[width=4.5cm]{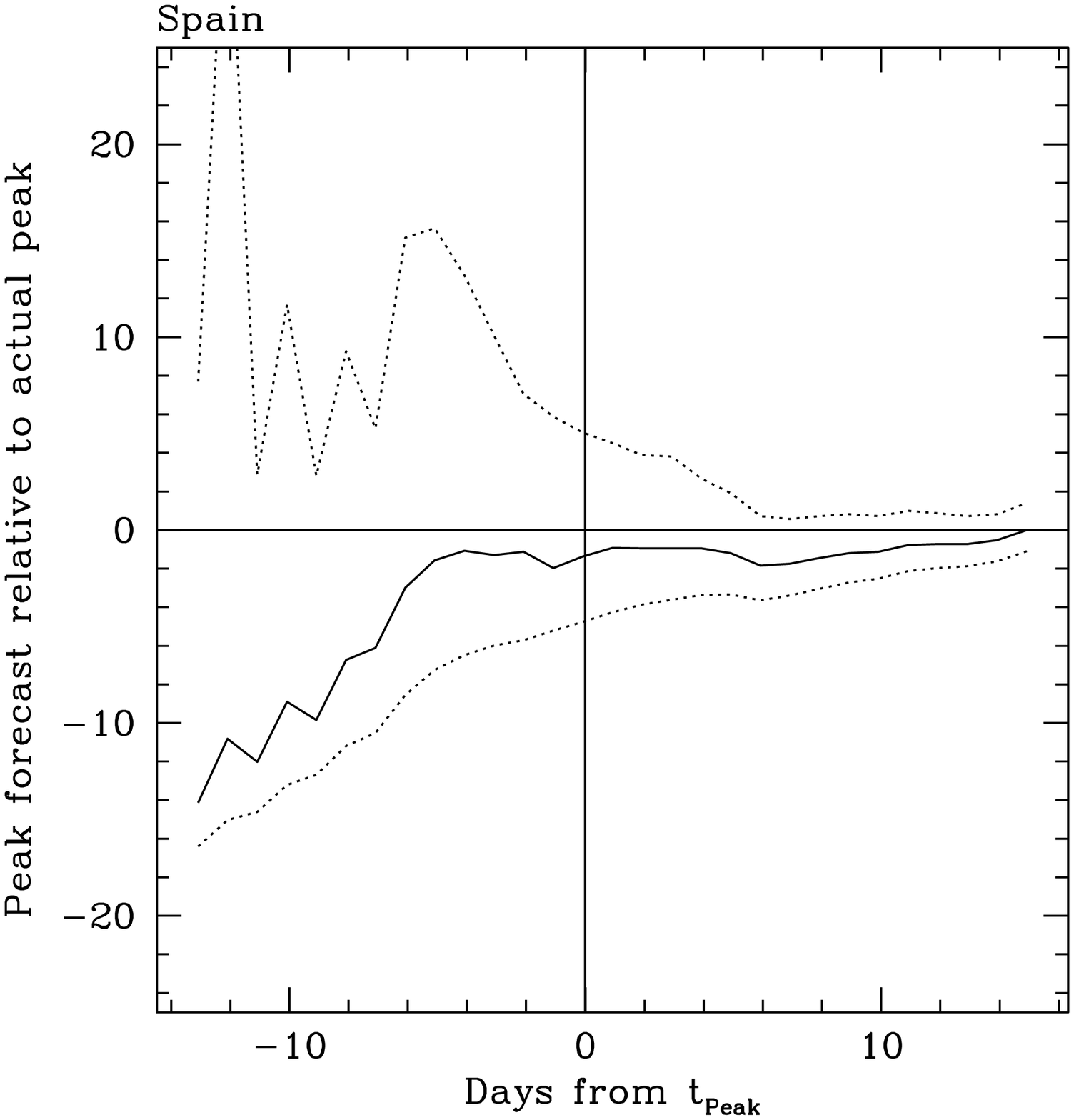}
\\
\includegraphics[width=4.5cm]{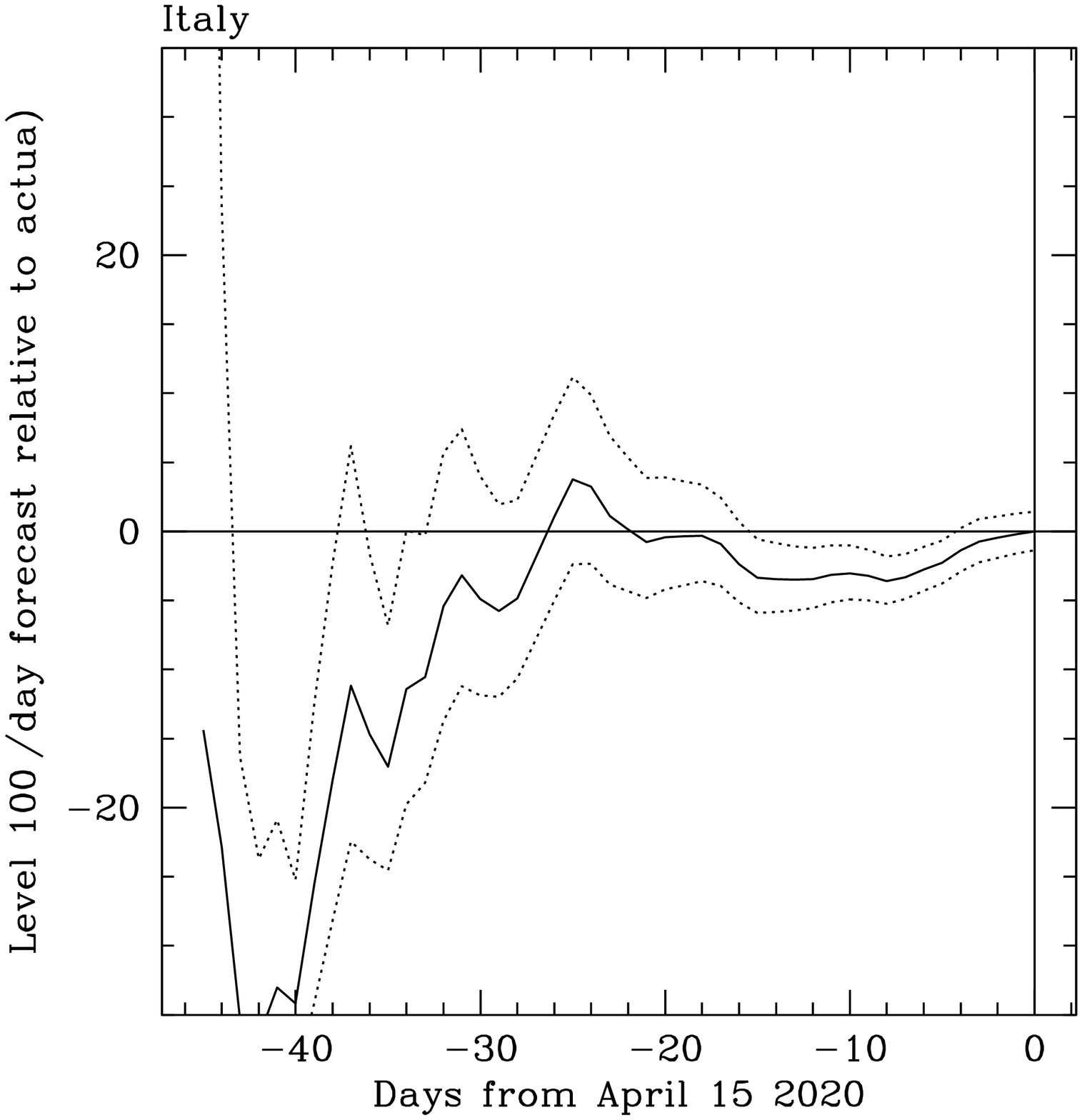}
\includegraphics[width=4.5cm]{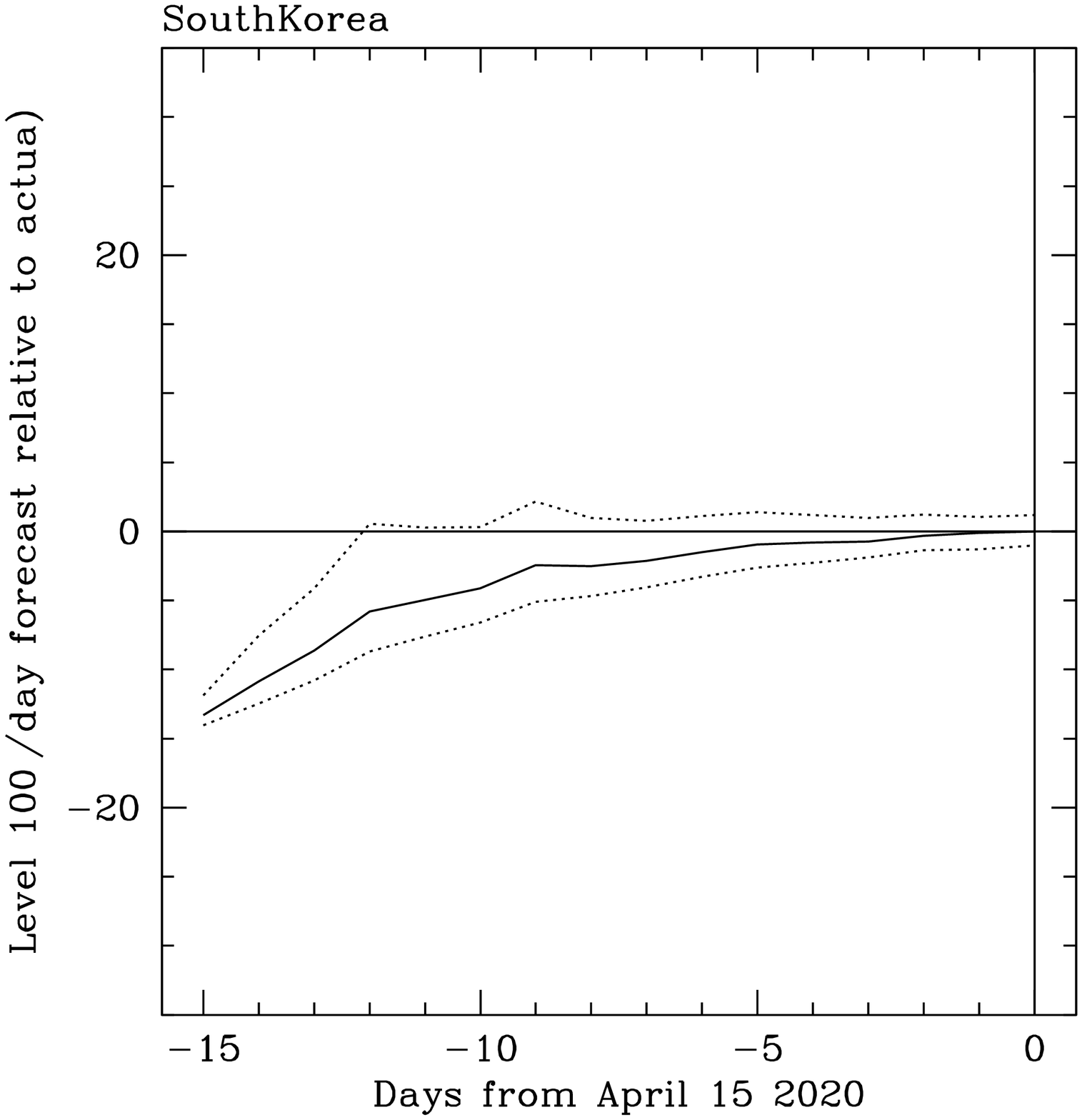}
\includegraphics[width=4.5cm]{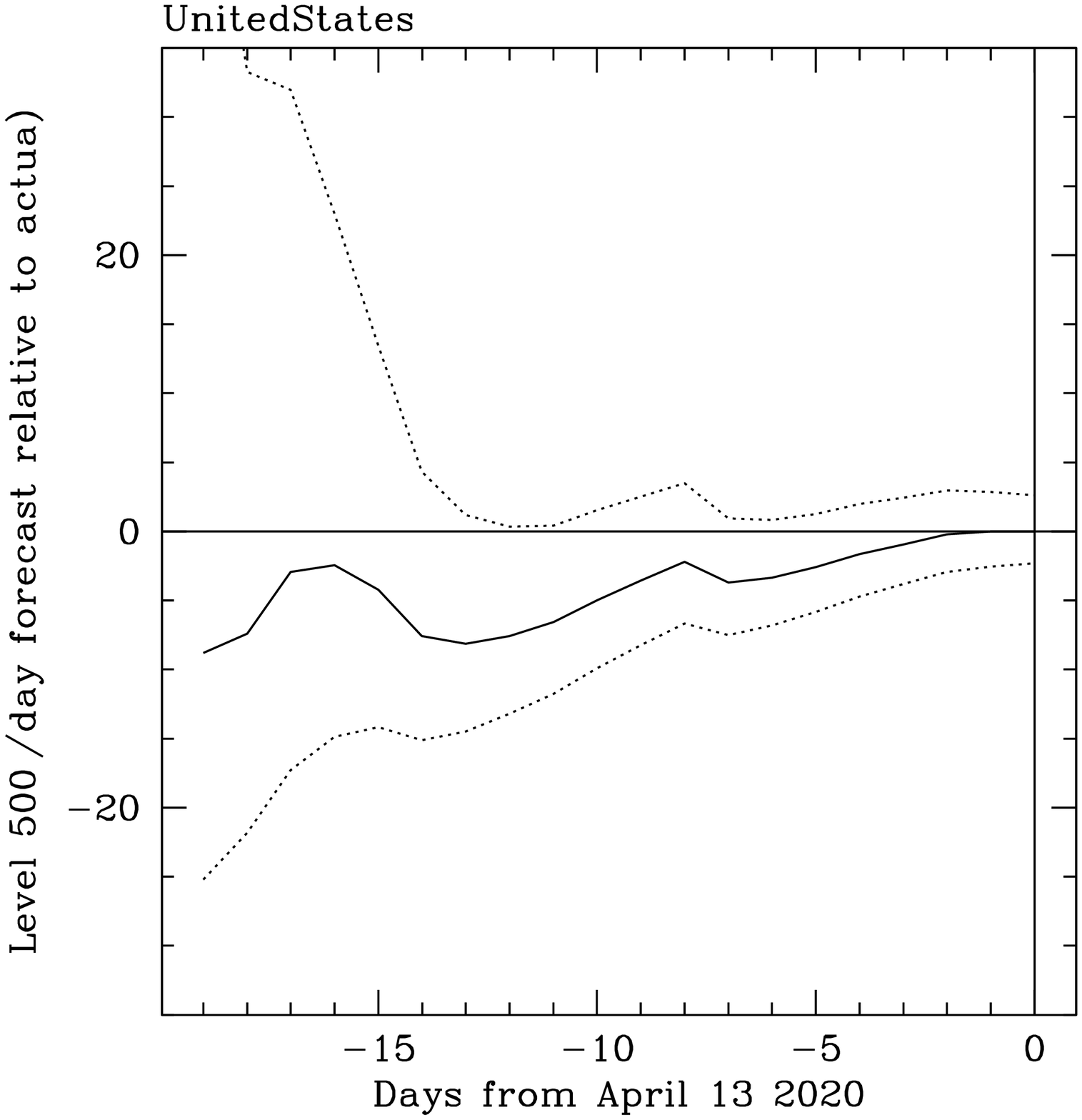}
\includegraphics[width=4.5cm]{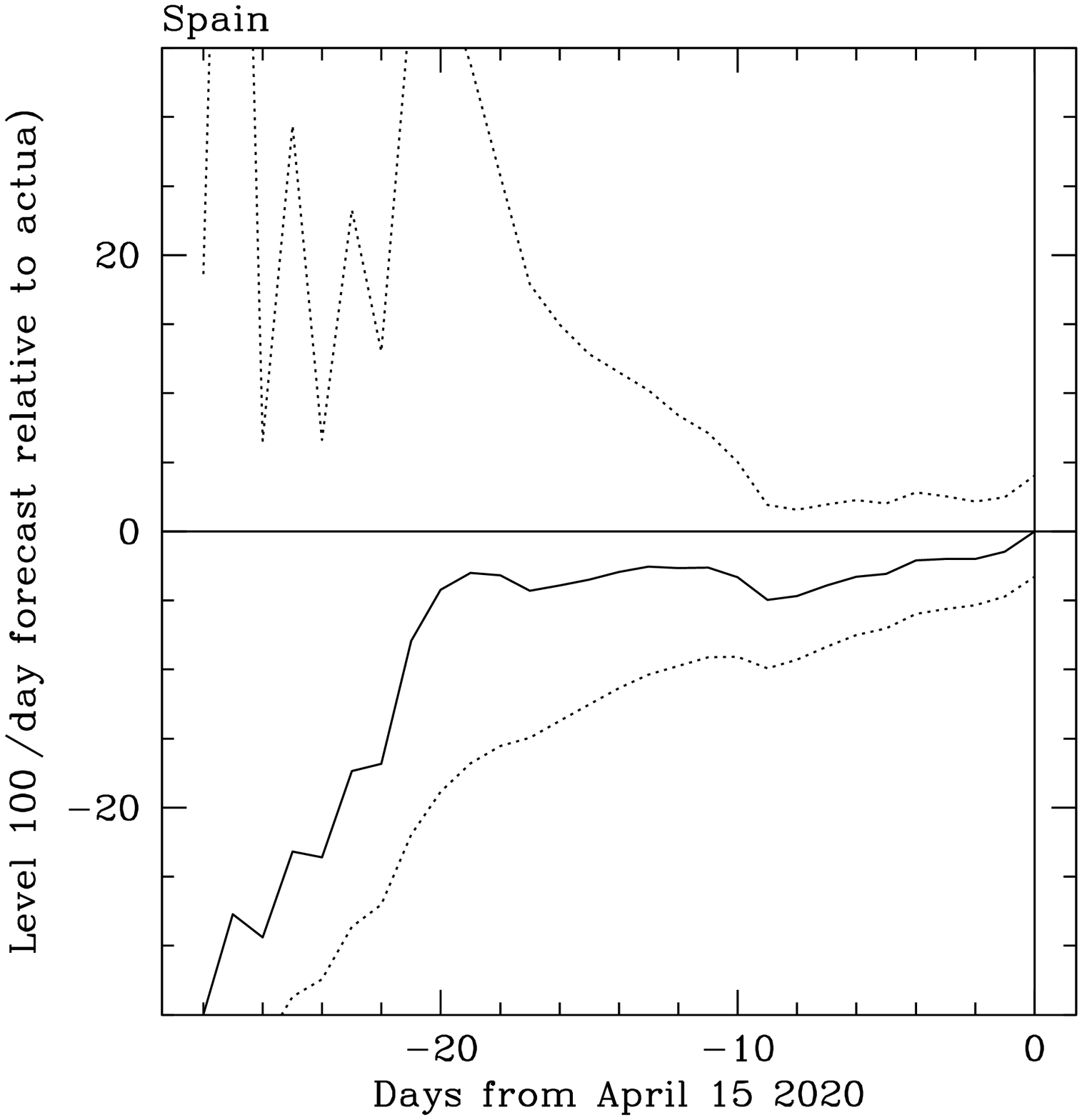}
\caption{\label{fig:peak}  {\bf Top panels:} Forecasting the peak in $\Delta N$ as a function of time for different countries, as labeled. Solid line shows best prediction and dotted line its 1$\sigma$ range, at each day. For Italy the peak forecast was accurate within 5~days, 20 days in advance of the peak, and about 10 days for the US and Spain. The peak was predicted only a few days in advance for South Korea, that reached the peak very rapidly.  {\bf Bottom panels:} Forecasting when various countries (as labeled) will reach a production rate of new infected per day ($\Delta N$) at the level of 2 per milion abitant, post peak. Lines are as in the top panels. For Italy accurate prediction of that regime is obtained with 30 days of advance, about 10-15 days for the US and Spain and 13 days for South Korea. We emphasize that as of the time of writing only South Korea has actually reached this level, and was able to maintain it for at least one month. 
}
\end{centering}
\end{figure*}

\section{The meaning and determination of Gumbel parameters}

Once the best-fit parameters $\mu$ and $a$ have been determined from the data by a fitting procedure, the Gumbel function provides an excellent 
description of the past time evolution of the COVID-19 disease and allows us to make predictions of the future evolution of the outbreak, including 
the peak in the number of new infections, as it is illustrated for the predicted evolution of $\Delta\, N$ with time in the case of Italy 
(Fig.~\ref{fig:ita1}-right). The parameter $a$, which has the physical dimension of time (we have chosen here the day as unit), sets the overall shape 
of the function, including the width of the peak where a larger value of $a$ implies a longer turnaround time for the outbreak. In particular, we find that 
the FWHM (Full Width at Half Maximum) of the peak can be approximated as 

\begin{equation}
{\rm FWHM}_{\rm Peak}\sim 0.63\times a+4.3.
\end{equation}

The parameter $\mu$ sets the timescale of the outbreak and thus allows one to predict the time (day) of the peak in $\Delta N$ ($t_{\rm peak}$, as we 
shall see below). Finally, there also is an overall normalization coefficient that can be set to reproduce the total number of infected people at the time 
of the peak, $N(t_{\rm peak})$. Notice that  the time of the peak can be  derived from the equation $e^x+x=ln(a)$, where $x=(t_{\rm peak}-\mu)/a$, 
which has the solution 

\begin{equation}
t_{\rm peak}=\mu+a(ln(a)-W(a)) 
\end{equation}

\noindent
where $W$ is the Lambert function. In Table~\ref{tab:table1} we report $t_{\rm peak}$, instead of $\mu$, since they are effectively equivalent 
parameters but the former has more immediate utility than the latter. Also notice that Eq.~\ref{eq:gumbel} cannot be analytically inverted to solve for 
$dN/dt$ and $N$,  and thus we have proceeded to numerically inverting it, integrating over the natural reporting timescale $\Delta t=1$~day.

It's interesting to emphasize that in the limit of a purely exponential growth, $N/\Delta N = \tau$ is constant, corresponding to a Gumbel function with $a$ that tends to $\infty$ and with exp(exp($-a\mu$))$=\tau$.
The exponential growth is thus a special case in the Gumbel growth we are describing, with infinite timescale $a$.

\section{Verifying the accuracy of the model's predictions}
\label{sec:verify}

If the future evolution continues along the same trend of the best-fit function, i.e. with the same functional form and identical parameters, as the data 
suggest, the Gumbel function can be used to predict the near future evolution of the outbreak, including providing an estimate of the uncertainty. 
Such a prediction is shown for the Italian data in Fig.~\ref{fig:ita1}-right. Similar predictions for South Korea, Spain, France and the US are also 
shown in Fig.~\ref{fig:more}. At 
the time of this writing, Italy, Spain, France and the US are the countries with the largest number of reported cases. As the plots show (data updated 
as of April 13), according to our analysis, all of them have already (cumulatively, not necessarily on a region-by-region case) passed the peak of the daily new 
infected ($\Delta\, N$). South Korea has actually reached a steady state of daily new infection with $\Delta\, N\approx 50$--100 (not included in the 
plots shown; see {\it worldometer}), thus demonstrating that such a low level can be kept constant over several weeks 
at least, preventing further exponential diffusion of the infection. 

The excellent fit provided by the Gumbel function to the time dependence of $\Delta\, N/N$, with constant parameters $a$ and $\mu$ over a large 
time interval  (Eqs.~\ref{eq:gumbel}), as shown in left panels of Figures~\ref{fig:ita1}--\ref{fig:regions}, suggests that, with 
adequate sampling, the same best fitting parameters could have been determined significantly earlier on during the diffusion of the COVID-19 
infection in each country. If that could indeed be made possible one would have been able to predict key properties of the evolution of the outbreak, 
such as the timing of the peak in $\Delta\, N$ and its width, the point in time when $\Delta\, N$ reaches a pre-set value determined to allow a country 
to maintain a manageable steady state control of the disease, for example similar to what achieved by South Korea.

To further investigate the feasibility of such predictions in the countries considered here, we have performed the following simulation: we 
have fit the time series of $\Delta\,N /N$ only up to a given day in the past, ignoring the additional data points after that, and compared the prediction 
obtained in this way with the one obtained using the full data set available at this time, including a detailed analysis of the relative uncertainties. 
Fig.~\ref{fig:peak} top-panels shows the results of this simulation to illustrate the ability of our methodology to predict the peak in $\Delta N$. The graphs 
clearly show that the peak could have been predicted as much ahead in time as about 3 weeks for Italy and 1-2~weeks for the other countries. For 
most values of the time lapse that we have considered, a reasonable 1$\sigma$ uncertainty for the peak is also predicted, which in fact brackets the 
true date of the peak. For Italy and South Korea a small underestimate of the peak time is observed using the earliest data points. We also observe 
a similar behaviour in other countries that we have inspected, a fact that we interprete as due to the progressive settling of the reporting mechanisms 
and procedures for testing adopted by the various governments rather than a shortcoming of our model.

The bottom panels of Fig.\ref{fig:peak} show the prediction of the time, after the peak, when the epidemic outbreak can be considered to be under 
control:  in South Korea that time was when it reached $\Delta\, N\sim100$ per day, a level that has been so far maintained for 
about 1 month without the occurrence of a second outbreak wave. This corresponds to approximately 2 new infected per day per million people.  
Using our model, such crucial moment in time  can be predicted with large anticipation, up to about 35 days in the case of the Italian data. Of course, 
this assumes that the evolution continues with the same Gumbel parameters down to this {\it floor}: so far, only in the case of South Korea 
such a level  has been convincingly reached. 

\begin{figure}[ht]
\begin{centering}
\includegraphics[width=8cm]{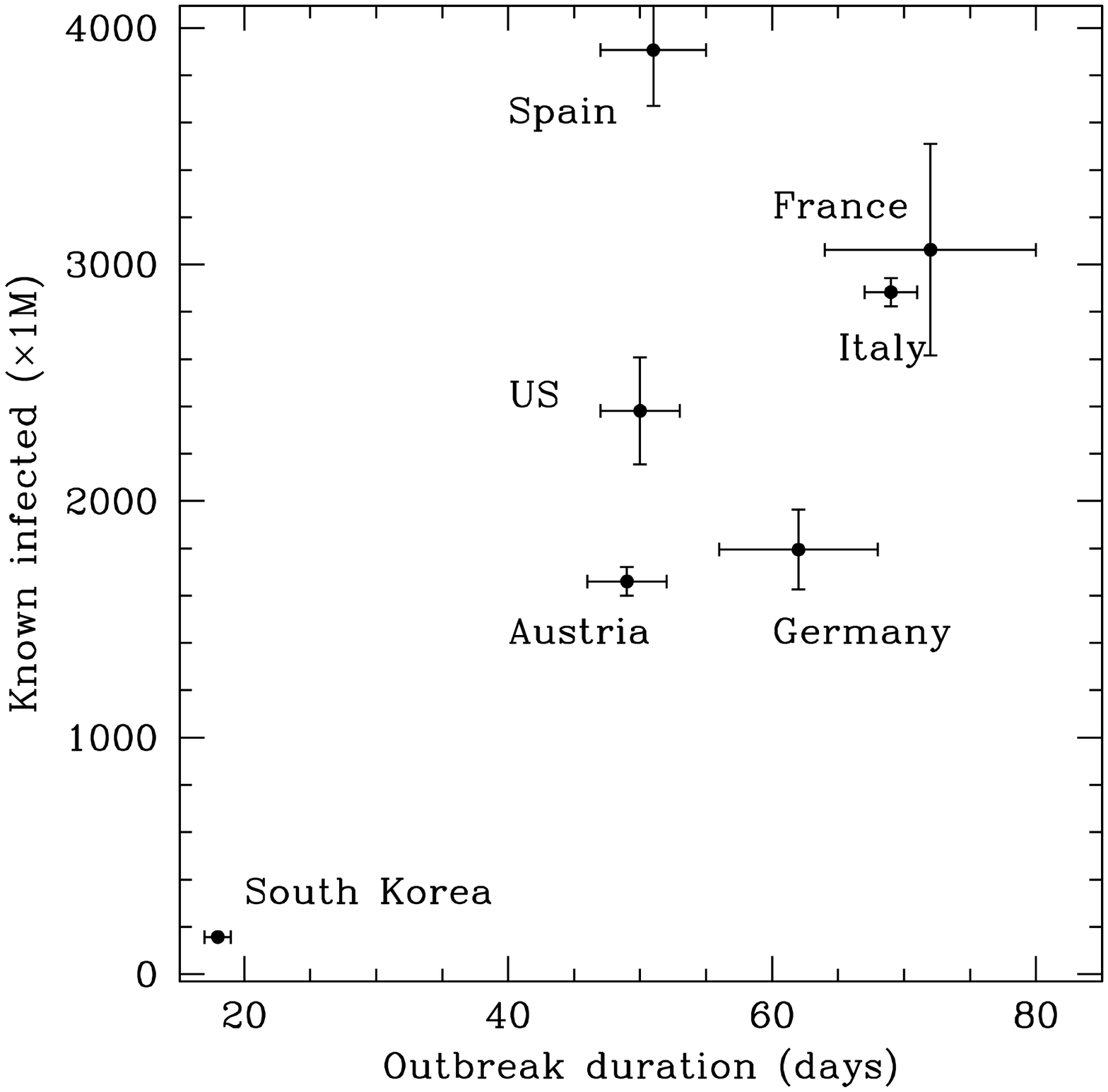}
\caption{\label{fig:compare}  We compare performances in countering the break from various countries. The x-axis shows the duration of the outbreak, defined as the time during which the number of new infected ($\Delta N$) was higher than about 2/day/M. The y-axis shows the total (at the end of the wave, not just until observed) number of known infected people in the outbreak.  Among this four, Italy and France have been the poorest performers in terms of speed in countering the outbreak, probably evidence of lower efficiency in the countermeasures. Spain reached the highest number of infected people per 1M population, suggesting they started countering the virus late. South Korea did outperform the other countries by over one order of magnitude in terms of number of infected per million, and by a factor of 3-4 in terms of duration.
}
\end{centering}
\end{figure}

\begin{figure}[ht]
\begin{centering}
\includegraphics[width=8cm]{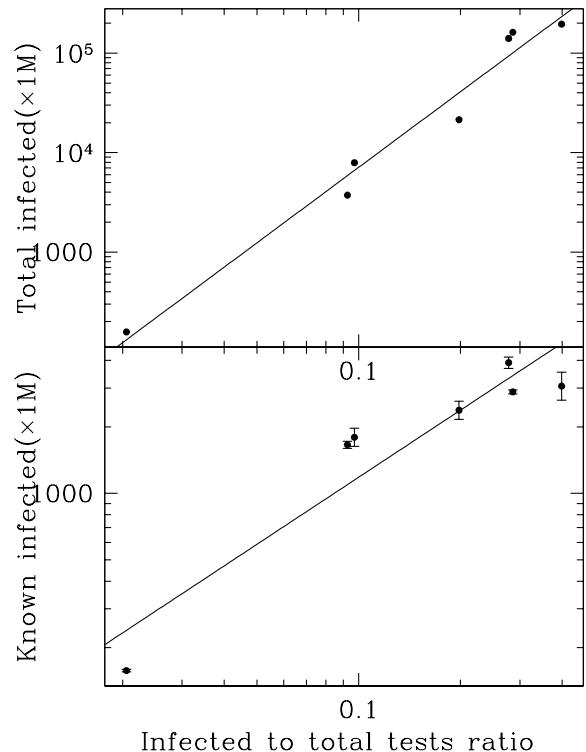}
\caption{\label{fig:xcorr}  {\bf Bottom:} Correlation of infected to total test ratios versus total number of infected per milion in selected countries (from left to right points are from South Korea, Austria, Germany, US, Spain, Lombardia and France -- we use Lombardia in representation of Italy as it carries the vast majority of its infections). {\bf Top:} Correlation with total infected per 1M, with a correction attempting to account for both the undercounting of known infected and known deaths, assuming a universal CFR for COVID-19.
}
\end{centering}
\end{figure}

\begin{figure}[ht]
\begin{centering}
\includegraphics[width=8cm]{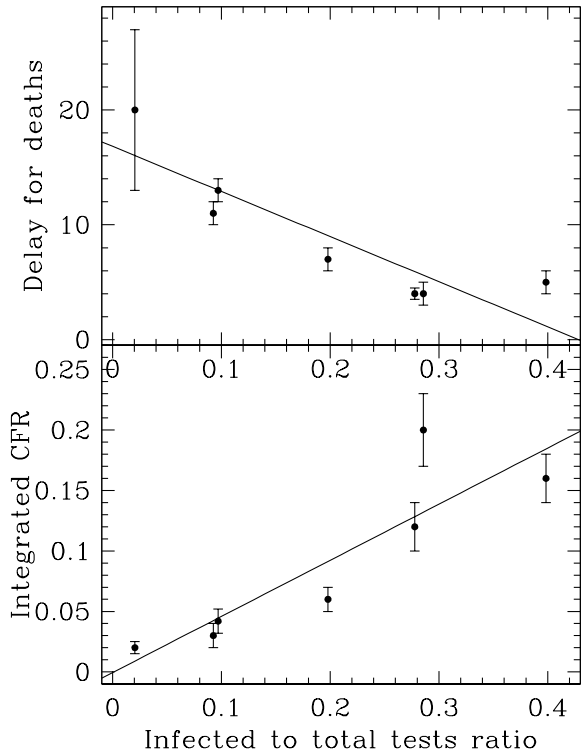}
\caption{\label{fig:xcorr2}  Correlation of infected to total test ratios versus the integrated CFR (bottom) and the delay between infected and deaths (top), for the same countries as in Fig.~\ref{fig:xcorr}. The delay is apparent from Figs~1--3 and is measured as the difference in the $\mu$ parameters from the Gumbel fit. 
}
\end{centering}
\end{figure}

\begin{table*}[ht]
\caption{\label{tab:table1}%
We report various estimates from our Gumbel modeling, using data collected until  April 26th 2020.
}
\begin{ruledtabular}
\begin{tabular}{lc|ccccccccccc}
\textrm{Country/}& \textrm{Pop.}& \multicolumn{2}{c}{\textrm{N final}} & \multicolumn{2}{c}{\textrm{$\Delta N$ Peak}}  &  Peak & Floor\footnote{This is defined as reaching the level of 2/day/M new infections} &  Duration & a 
 & rms\footnote{This refer to the fit of log$_e$(log$_e$($N/\Delta N$)) as in Figs~\ref{fig:ita1}--\ref{fig:regions} (which show log$_{10}$ instead -- the $rms$ in double log$_{10}$ space is 0.189 times the one reported here)} & $PCC$\footnote{Pearson  correlation coefficient between log(log($N/\Delta N$) and time} \\
 Region & 10$^6$ & Total & per 10$^6$ & Total & per 10$^6$ & & & days & days & & \\
\colrule
S. Korea & 51.4 & $ 8093_{-58}^{+110} $ & $ 157_{-1}^{+2} $ & $ 690_{-39}^{+37} $ & $ 13_{-0.7}^{+0.7} $ & {\rm March} $ 1_{-1}^{+1} $ & {\rm March} $ 11_{-1}^{+2} $ & $ 19_{-1}^{+2} $ & $ 11.0\pm1.0 $ & 0.23 & 0.93\\
Austria &  9.3 & $ 15341_{-19}^{+27} $ & $ 1658_{-2}^{+2} $ & $ 611_{-17}^{+17} $ & $ 66_{-1}^{+1} $ & {\rm March} $ 29_{-1}^{+1} $ & {\rm April} $ 27_{-1}^{+2} $ & $ 54_{-2}^{+2} $ & $ 30.0\pm 1.1 $ & 0.14 & 0.96 \\   
Switzerland &  8.6 & $ 29424_{-94}^{+142} $ & $ 3427_{-10}^{+16} $ & $ 1115_{-45}^{+44} $ & $ 129_{-5}^{+5} $ & {\rm March} $ 30_{-1}^{+1} $ & {\rm May} $ 3_{-1}^{+3} $ & $ 58_{-2}^{+3} $ & $ 32.0\pm 1.8 $ & 0.19 & 0.93 \\ 
Germany & 83.0 & $ 162446_{-1606}^{+2246} $ & $ 1958_{-19}^{+27} $ & $ 5505_{-185}^{+173} $ & $ 66_{-2}^{+2} $ & {\rm April} $ 1_{-1}^{+2} $ & {\rm May} $ 6_{-1}^{+3} $ & $ 63_{-3}^{+3} $ & $ 36.9\pm 2.1 $ & 0.17 & 0.93 \\ 
France & 67.1 & $ 173209_{-5002}^{+8889} $ & $ 2580_{-74}^{+132} $ & $ 5678_{-302}^{+227} $ & $ 84_{-4}^{+3} $ & {\rm April} $ 5_{-2}^{+3} $ & {\rm May} $ 13_{-3}^{+6} $ & $ 68_{-5}^{+7} $ & $ 38.5\pm 4.2 $ & 0.28 & 0.80 \\    
Spain & 46.3 & $ 238843_{-3114}^{+4560} $ & $ 5163_{-67}^{+98} $ & $ 8138_{-343}^{+317} $ & $ 175_{-7}^{+6} $ & {\rm April} $ 3_{-1}^{+2} $ & {\rm May} $ 13_{-1}^{+4} $ & $ 64_{-3}^{+4} $ & $ 36.7\pm 2.6 $ & 0.18 & 0.90 \\
Italy & 60.7 & $ 208423_{-1249}^{+1439} $ & $ 3436_{-20}^{+23} $ & $ 5960_{-79}^{+76} $ & $ 98_{-1}^{+1} $ & {\rm March} $ 31_{-1}^{+1} $ & {\rm May} $ 14_{-1}^{+2} $ & $ 79_{-2}^{+2} $ & $ 45.8\pm 1.1 $ & 0.08 & 0.98 \\     
Turkey & 82.0 & $ 127797_{-4061}^{+5336} $ & $ 1558_{-49}^{+65} $ & $ 4704_{-71}^{+44} $ & $ 57_{-1}^{+1} $ & {\rm April} $ 12_{-1}^{+2} $ & {\rm May} $ 14_{-1}^{+3} $ & $ 53_{-3}^{+3} $ & $ 33.2\pm 2.1 $ & 0.11 & 0.94 \\ 
Iran & 81.8 & $ 101074_{-1835}^{+2277} $ & $ 1235_{-22}^{+27} $ & $ 2332_{-50}^{+45} $ & $ 28_{-1}^{+1} $ & {\rm March} $ 31_{-1}^{+2} $ & {\rm May} $ 16_{-1}^{+3} $ & $ 74_{-3}^{+3} $ & $ 60.0\pm 3.1 $ & 0.10 & 0.94 \\
Chile & 18.7 & $ 16456_{-897}^{+1303} $ & $ 879_{-47}^{+69} $ & $ 487_{-5}^{+1} $ & $ 26_{-1}^{+1} $ & {\rm April} $ 12_{-2}^{+3} $ & {\rm May} $ 17_{-2}^{+5} $ & $ 61_{-4}^{+5} $ & $ 43.9\pm 3.9 $ & 0.14 & 0.88 \\ 
Netherlands & 17.3 & $ 42172_{-811}^{+1020} $ & $ 2441_{-46}^{+59} $ & $ 1288_{-21}^{+17} $ & $ 74_{-1}^{+1} $ & {\rm April} $ 7_{-1}^{+2} $ & {\rm May} $ 17_{-1}^{+3} $ & $ 71_{-3}^{+3} $ & $ 42.1\pm 2.0 $ & 0.11 & 0.95 \\   
US & 330.5 & $ 1.12M_{-24138}^{+29624} $ & $ 3387_{-73}^{+89} $ & $ 38730_{-473}^{+363} $ & $ 117_{-1}^{+1} $ & {\rm April} $ 10_{-1}^{+1} $ & {\rm May} $ 18_{-1}^{+3} $ & $ 63_{-3}^{+3} $ & $ 36.0\pm 1.6 $ & 0.09 & 0.96 \\
UK & 67.1 & $ 192158_{-8443}^{+10998} $ & $ 2862_{-125}^{+163} $ & $ 5918_{-7}^{+38} $ & $ 88_{-1}^{+1} $ & {\rm April} $ 13_{-2}^{+2} $ & {\rm May} $ 24_{-2}^{+4} $ & $ 69_{-4}^{+4} $ & $ 41.7\pm 2.5 $ & 0.11 & 0.94 \\
Sweden & 10.2 & $ 25096_{-1947}^{+2972} $ & $ 2453_{-190}^{+290} $ & $ 569_{-2}^{+16} $ & $ 55_{-1}^{+1} $ & {\rm April} $ 12_{-3}^{+4} $ & {\rm June} $ 4_{-5}^{+9} $ & $ 90_{-7}^{+9} $ & $ 61.4\pm 6.3 $ & 0.18 & 0.81 \\  
Peru & 32.0 & $ 61943_{-17763}^{+50885} $ & $ 1937_{-555}^{+1591} $ & $ 1623_{-257}^{+855} $ & $ 50_{-8}^{+26} $ & {\rm April} $ 27_{-7}^{+11} $ & {\rm June} $ 10_{-13}^{+5} $ & $ 77_{-15}^{+5} $ & $ 51.3\pm12.2 $ & 0.23 & 0.61 \\
Russia & 144.5 & $ 482k_{-133k}^{+255k} $ & $ 3338_{-921}^{+1767} $ & $ 9825_{-2034}^{+3730} $ & $ 68_{-14}^{+25} $ & {\rm May} $ 11_{-6}^{+7} $ & {\rm July} $ 12_{-11}^{+18} $ & $ 104_{-13}^{+18} $ & $ 70.2\pm 8.4 $ & 0.07 & 0.85 \\ 
SaudiAr. & 33.7 & $ 185987_{-112k}^{+1.2M} $ & $ 5515_{-3349}^{+37153} $ & $ 2514_{-1260}^{+11277} $ & $ 74_{-37}^{+334} $ & {\rm May} $ 28_{-20}^{+44} $ & {\rm Aug.} $ 31_{-47}^{+118} $ & $ 162_{-48}^{+117} $ & $ 116.5\pm39.9 $ & 0.19 & 0.44 \\  
\hline
Sicilia &  5.1 & $ 3236_{-51}^{+77} $ & $ 629_{-10}^{+15} $ & $ 99_{-4}^{+3} $ & $ 19_{-1}^{+1} $ & {\rm April} $ 2_{-2}^{+2} $ & {\rm May} $ 3_{-1}^{+4} $ & $ 57_{-3}^{+4} $ & $ 42.0\pm 3.3 $ & 0.20 & 0.88 \\      
Campania &  5.8 & $ 4530_{-67}^{+105} $ & $ 780_{-11}^{+18} $ & $ 137_{-6}^{+5} $ & $ 23_{-1}^{+1} $ & {\rm March} $ 31_{-2}^{+2} $ & {\rm May} $ 3_{-2}^{+4} $ & $ 59_{-4}^{+4} $ & $ 42.5\pm 3.4 $ & 0.23 & 0.86 \\      
Marche &  1.5 & $ 6187_{-18}^{+26} $ & $ 4150_{-12}^{+17} $ & $ 219_{-7}^{+7} $ & $ 147_{-4}^{+4} $ & {\rm March} $ 29_{-1}^{+1} $ & {\rm May} $ 5_{-1}^{+3} $ & $ 64_{-2}^{+3} $ & $ 34.8\pm 1.6 $ & 0.17 & 0.94 \\
Lazio &  5.8 & $ 6667_{-76}^{+103} $ & $ 1158_{-13}^{+18} $ & $ 214_{-6}^{+6} $ & $ 37_{-1}^{+1} $ & {\rm April} $ 2_{-1}^{+2} $ & {\rm May} $ 6_{-1}^{+3} $ & $ 61_{-3}^{+3} $ & $ 39.4\pm 2.2 $ & 0.16 & 0.93 \\ 
Toscana &  3.7 & $ 9447_{-79}^{+109} $ & $ 2552_{-21}^{+29} $ & $ 314_{-10}^{+9} $ & $ 85_{-2}^{+2} $ & {\rm April} $ 1_{-1}^{+1} $ & {\rm May} $ 8_{-1}^{+3} $ & $ 66_{-3}^{+3} $ & $ 37.7\pm 2.0 $ & 0.17 & 0.93 \\
Emilia Rom. &  4.6 & $ 25395_{-191}^{+254} $ & $ 5551_{-41}^{+55} $ & $ 782_{-21}^{+20} $ & $ 170_{-4}^{+4} $ & {\rm March} $ 30_{-1}^{+1} $ & {\rm May} $ 13_{-1}^{+3} $ & $ 78_{-3}^{+3} $ & $ 41.7\pm 1.9 $ & 0.15 & 0.94 \\
Veneto &  4.9 & $ 18630_{-301}^{+417} $ & $ 3815_{-61}^{+85} $ & $ 502_{-15}^{+14} $ & $ 102_{-3}^{+2} $ & {\rm March} $ 31_{-2}^{+2} $ & {\rm May} $ 18_{-2}^{+4} $ & $ 83_{-4}^{+4} $ & $ 49.3\pm 3.1 $ & 0.17 & 0.90 \\
Lombardia & 10.0 & $ 76855_{-933}^{+1271} $ & $ 7667_{-93}^{+126} $ & $ 2115_{-67}^{+62} $ & $ 211_{-6}^{+6} $ & {\rm March} $ 30_{-2}^{+2} $ & {\rm May} $ 20_{-2}^{+4} $ & $ 85_{-4}^{+4} $ & $ 48.1\pm 2.7 $ & 0.16 & 0.92 \\
Piemonte &  4.4 & $ 27433_{-1070}^{+1950} $ & $ 6279_{-244}^{+446} $ & $ 809_{-38}^{+22} $ & $ 185_{-8}^{+5} $ & {\rm April} $ 5_{-3}^{+4} $ & {\rm May} $ 23_{-5}^{+9} $ & $ 82_{-7}^{+9} $ & $ 44.0\pm 5.1 $ & 0.32 & 0.76 \\ 
\end{tabular}
\end{ruledtabular}
\end{table*}

\begin{table}[h]
\caption{\label{tab:table2}%
Estimates of the outbreak diffusion into the population}
\begin{ruledtabular}
\begin{tabular}{lccc}
\textrm{Country}&  \multicolumn{3}{c}{\textrm{Fraction of  population infected (percent)}}   \\
 & naive & corrected by CFR & twice corrected\footnote{See Section~\ref{sec:cfr}} \\
\colrule
      South Korea &    0.016  &   0.016  &   0.06\\
     Austria  &     0.17 & 0.25   &   0.37\\
     Germany    &  0.18   &   0.38  &    0.79\\
          US  &    0.24  &    0.71    &   2.1\\
       Spain   &   0.39   &    2.3  &     14.0 \\
       Italy  &    0.29   &    2.2    &   16.2\\
      France   &   0.31  &      2.5     &  19.8
\end{tabular}
\end{ruledtabular}
\end{table}

\section{Quantitative assessment of countries performances in countering COVID-19}

Our quantitative analysis of the evolution of the COVID-19 in various countries allows us to measure their efficiency in countering the virus spread 
from various parameters. This is based for simplicity in the countries with the largest number of cases, as a reference.
We report key quantities for a wider list of  countries and regions in Table~\ref{tab:table1}, so that the reader might extend our considerations elsewhere, if interested.

\subsection{Rapidity in halting and suppressing the infection spread} 

The parameter $a$, the slope of the double log of $\Delta N/N$, is the most important parameter because it sets the duration of the outbreak: the 
{\it Duration} of the $\Delta N$ curve defined as the number of days during which more than 2/day/M new infected are produced. We find that this 
{\it Duration} can be estimated as $1.75\times a$ (Table~\ref{tab:table1}). 

Figure~\ref{fig:compare} compares the performance of various countries in containing the disease: in Spain, the US and Austria the virus countering 
measures seem to be  more effective than Italy and Germany, since the former countries have shorter outbreaks and more the reproduction rate of 
the virus declines faster.  By far, South Korea has been over-performing all other countries in terms for its rapidity in suppressing the outbreak.

\subsection{Timeliness of the response} 

The total number of people infected in a country  ($N_{\rm Total}$), depends on the rapidity in containing the spread of the virus (larger $a$ brings 
more total infections) but also on how early measures are taken, i.e. the rapidity in changing the behaviour of $N/\Delta\, N$ from almost constant 
(exponential growth) into a Gumbel-like deceleration. The following formulation is a very good description of our data: 
$N_{\rm Total} = \Delta N_{\rm Peak}\times(0.66a+4.8)$. However, $\Delta N_{\rm Peak}$ grows linearly with $a$ at fixed normalization, implying that 
$N_{\rm Total}$ goes like $a^2$. However, also the normalization increases $N_{\rm Total}$ linearly. 
Fig.~\ref{fig:compare} shows that at fixed outbreak duration, Spain is producing twice the number of infected people per million than Austria and the 
US: we suggest that the worse performance is due to the longer delay before substantial containment measures have been adopted. Similarly, Italy 
and France appear to have responded later than Germany. 

\subsection{Reliability in reporting}

The rms scatter of the residual from the Gumbel fit to the $N/\Delta\, NN$ time series, and/or the Pearson correlation coefficient, can be 
related to the day-to-day consistency and reliability in reporting data (Table~\ref{tab:table1}). The best performing countries appear to be Italy and the 
United States, while France appears to be the least well performing.  South Korea also did not appear to provide very careful day-to-day reporting, which is also very 
well illustrated by time series of deaths $D/\Delta\, D$ that appears to be roughly constant with time and impossible to correlate with the corresponding 
data for the infections, after accounting for an appropriate delay time. Hence we set the delay to zero for display purposes in Fig.3, but estimate the delay to be actually quite long, as demonstrated that deaths are still counted in significant numbers to today (see {\it worldometers}).

\subsection{Relation to testing rates} 
\label{sec:cfr}

Different values of the ratio of infected to total tests are observed, too. When a large fraction of tested patients is found to be positive for the COVID-19 
infection, one can argue not only that a substantial fraction of infected people are not censed, but also that in general the quality of the virus countering 
efforts are inadequate. This is illustrated by Figure~\ref{fig:xcorr}, where we show the correlation (Pearson correlation coefficient 0.94) between 
the number of infections per million versus the fraction of positive tests. Similar levels of correlation are also observed with the parameter $a$ 
(not surprisingly, given Fig.~\ref{fig:compare}) and, more interestingly, with the integrated case fatality rate (CFR), see Fig.~7-bottom. We derive the CFR in each 
country in which the outbreak is still actively occurring by taking into account the delay between the fractional growth rate of infected and deaths. In fact, 
an excellent correlation is observed for CFR with this time delay between $N\Delta N$ of infected and $D/\Delta D$ of deaths (Fig.~7-top). This is obvious when considering that larger testing rates will discover larger numbers of less severely infected patients on average, and with the COVID-19 disease at an earlier stage. 

Assuming that the intrinsic fatality of the virus is the same in all countries, Fig.~\ref{fig:xcorr2} suggests that we can correct for the unaccounted 
infections by normalizing each country to the same CFR of 2\% as observed in South Korea, and consistent to what seen in the Diamond Princess 
ship outbreak (Russell et al 2020; [27]), where everyone of the 3711 passengers were tested. Using this procedure, we derive 
the numbers listed in Table~II (see CFR-corrected column). This suggests that when the current outbreak waves have subsided, up to about 
1-2\% of the population of several countries will have been infected, quite larger than the naive estimates of $<0.3$\% from the reported infections. 

This might still largely underestimate the actual number of infections in a country: when a substantial fraction of COVID-19 patients remain untested, 
it is reasonable to believe that also deaths are undercounted as well. Estimates based on registries of deaths in various municipalities in Italy have 
shown [24; 25; 26] that this deaths under-reporting factor can reach up to 4--6$\times$. We make the ad-hoc assumption that this further correction 
scales with the fraction of positive tests in exactly the same way as for the previous one (ratio of observed CFR to an intrinsic CFR of 2\%). We are 
of course aware that this calculation is affected by uncertainty and a reliable assessment will require a widespread program of population testing to 
search for the prevalence of antibody to the SARS-CoV-2 virus. It is interesting that the numbers suggest that up to 15-20\% 
of the population will infected (by the end of current outbreak waves) in Spain, Italy and France, a level not too far from the herd immunity (the numbers are  much lower in other countries). This implies that containment in these three
countries has been the least efficient.
It is intriguing that in the Diamond Princess ship, despite the very confined environment, which favored close contacts with infected passengers, only 
about 20\% of the passenger eventually tested positive, a fractions similar to the one inferred for Italy, Spain and France.  
\section{Discussion}

\subsection{A Universal description of outbreaks evolution}

A significant finding of this study is that, in presence of containment measures, the evolution with time of $\Delta N/N$, the fractional rate of new 
infections 
in an epidemic outbreak, is remarkably well modeled by the Gumbel function, i.e it has a universal shape regardless of the type of adopted measures. 
In other words, while the Gumbel parameters $a$ and $\mu$ (i.e., $t_{\rm Peak}$) vary from country to country, and from region to region within the 
same country (e.g. Italy), reflecting the diversity and degree of effectiveness of the adopted containment measures, the time evolution of new 
infections, which would be exponential in absence of any measure, always converges to the same functional relationship, i.e. the Gumbel function.

It is fundamental to notice that, as a result of the often highly inadequate and biased testing procedure adopted by each country, we do not actually 
have either complete or unbiassed measures of the true number of infected people (very often only people with severe symptoms of the COVID-19 
disease get tested for the virus, which bias the sample towards individual who most susceptible to the disease, and only relatively small numbers of 
tests are conducted). These systematic effects combine with the varying effectiveness of the containment measures adopted by the specific country 
(degree of enforcement of social distancing, number and types of allowed essential activities, availability of protective devices) to produce the 
"effective" metrics that are used in studies such as ours (in this case the variable $N(t)$). The fact that these "effective metrics" all obey the same 
functional relationship in such different societies as South Korea, Spain, Italy and the US, strongly suggests that this functional relationship is relatively
insensitive to all  systematics combined together. 
These systematics seem to always affect available metrics 
of the spread of the epidemic in the same way.  This is most likely the result of using the ratio of the rate of increase $\Delta\, N$ to $N$ itsef: 
even if $N$ is affected by a bias, which we can think of as a multiplicative factor, since the variable of relevance is actually $r=\Delta\, N/N$, to 
first order the bias is largely eliminated in the ratio. To us, this seems a non-trivial result, reminiscent of a situation where the combined effect of 
a large number of independent causes result in the well-defined, universal final behavior of a random variate (e.g. normal distribution and the Central 
Limit Theorem). In this case, $dr/r$, namely the daily fractional change of $r$, which is the same as $dR_0/R_0$, diverges exponentially with time towards more and more negative values (Eq.~3).

\subsection{Relation to the standard SIR model}

It is instructive to compare our model to the standard SIR model which is commonly used to describe epidemic outbreaks ([28-33]). In the SIR model a 
total population T of individuals experience an infectious epidemic is divided into $S$, $I$ and $R$ groups, namely are the numbers of susceptible, 
infected and recovered (or killed by the disease) individuals (notice that we have adopted the notation $N$ instead of $I$, that we keep in the following 
for clarity), where $T=S+N+R$ (all normalized to the total number of individuals to that $T$ so that $T=1$). The rates at 
which people are infected is regulated by the constants $\beta$ and $\gamma$, such that $dS/dt=-\beta\, S\, N$, $dN/dt=\beta\, S\, N-\gamma\, N$ 
and $dR/dt=\gamma\, N$.  In this model, which provides a description of an outbreak in absence of any measure aimed at slowing and reversing the 
spread of the disease, the quantities $\gamma$ and $\beta$ are constant in time. The parameter $R_0=\beta/\gamma$ ([34]), usually called "the 
reproduction number", which measures the  average number of secondary infected individual created by 1 primary infected primary, is often used 
as a metric of the 
effective power of the contagion in the outbreak. The early phase is characterized by $I=I_i\ll 1$ and $S_i=1-I_\approx 1$ and thus 

\begin{equation}
	N(t)=N_i\, e^{(R_0-1)\, \gamma\, t}.
\end{equation}

If $R_0>1$ one has an epidemic outbreak. If left unchecked, the epidemic will continue until $S$ substantially decreases and $N$ and $\Delta\, N$ 
eventually deviates from the exponential growth and $\Delta\, N$ , in particular, reaches a peak: as a result of the decreased efficiency of new 
infections and the decrease of $N(t)$ as a result of people who recover or die. Letting an epidemic evolve in this way would result in unacceptable 
loss of life and containment measures must be set in place to break the exponential growth during the early phases. Such containment measures 
basically aim at breaking the exponential growth during the early phase by making $R_0$ ($r$ in our model) become a function of time and bring 
it to $R_0\le 1$. It is important to realize that the peak that an unchecked epidemic would reach is qualitatively different from the peak that happens 
as a result of containment measures. The former basically happens because the population becomes depleted of Susceptible individuals; the latter 
is the direct result of the containment, which continuously shorten $R_0$, or $r$, which in effect is the inverse of the continuously increasing $\tau$ 
time-scale factor of the exponential growth (there is no exponential growth in an unchecked epidemic approaching the peak). 


In the previous section we have suggested that, even if the available diagnostics of an epidemic are biased (e.g. the available number of infected 
individuals is both an incomplete and biased estimator of the true value), a fortuitous property of how all the biases combine together makes the 
time evolution of the resulting "observed fractional growth rate", $\Delta\, N/N$, a universal function of time. In addition, the fact of using the ratio of 
the first derivative of a variable to the variable itself greatly diminishes the effect of any systematic bias. This universal function of time turns out to be 
very useful to predict both the time occurrence of the peak, its width and the time when the infection is basically under control. That said, it is 
nonetheless of considerable interest to estimate the total number of infections at a given time during the outbreak (as we have attempted in Section V~D), since this can be used to inform 
the testing strategy, optimize the containment measures, as well as to obtain a fair estimate the CFR of the disease, both cumulative and as a 
function of the age of the patients, useful to inform the clinical response. 


\section{Conclusions}

There are two main conclusions from this work, a conceptual one and a practical one. The conceptual one is that during the exponential phase of a 
pandemic outbreak and in presence of containment measures, the time evolution of the fractional growth rate of new infections, $\Delta\, N/N$,  
follows a universal functional form, which is very well modeled with the two-parameter Gumbel function. Fits to the data show that the two Gumbel 
parameters vary from country to country (and even from region to region within the same country), reflecting the specifics of the adopted containment 
measures. Remarkably, however, the functional form, remains the same. Since in absence of the containment measures, the evolution of the 
pandemic at this stage would still be in the exponential phase (as we have shown, the total number of infected is still small compared to the 
population (see Table~\ref{tab:table2}), it is possible that the containment measures modify the constant time scale of the exponential into a 
time-variable one, whose time-dependence is described by the Gumbel function. This is specified by two parameters $\mu$ and $a$, which we have 
fitted from the available data in a number of countries (Table~\ref{tab:table1}) and found that these parameters can be robustly constrained from the 
whole time span, or subsets of it, from the onset of the outbreak to the current time, and their value remain unchanged in each country. 

The main practical message of this work, which stems from the stability of the Gumbel parameters in each country, is that  reliable 
predictions of the future evolution of the COVID-19 outbreak, including key events, can be obtained with substantial advance to inform 
critical strategic decisions. These events include the time of the peak and the time when the rate of new infections reaches pre-set low level such 
that the epidemic can be managed and social and economic activities can be resumed. All that is required would be well-controlled and stable 
strategy of testing and measure of infections and tracking of the evolution of the daily fraction of new infections. This should be possible for most 
countries, provided that testing is carried out in a stable and well-controlled manner that ensures that the daily rate of new infections is robustly 
and consistently measured and reported. At the same time, such an analysis provides an effective monitor of the outbreak, helping governments 
to keep it at a manageable level by strengthening (or softening) the containment measures as needed. 

\begin{acknowledgments}
We wish to acknowledge enlightening discussions with many of our colleagues.
\end{acknowledgments}


\nocite{*}

\bibliography{ed_mg.bib}

\providecommand{\noopsort}[1]{}\providecommand{\singleletter}[1]{#1}%
\begin{thebibliography}{34}%
\makeatletter
\providecommand \@ifxundefined [1]{%
 \@ifx{#1\undefined}
}%
\providecommand \@ifnum [1]{%
 \ifnum #1\expandafter \@firstoftwo
 \else \expandafter \@secondoftwo
 \fi
}%
\providecommand \@ifx [1]{%
 \ifx #1\expandafter \@firstoftwo
 \else \expandafter \@secondoftwo
 \fi
}%
\providecommand \natexlab [1]{#1}%
\providecommand \enquote  [1]{``#1''}%
\providecommand \bibnamefont  [1]{#1}%
\providecommand \bibfnamefont [1]{#1}%
\providecommand \citenamefont [1]{#1}%
\providecommand \href@noop [0]{\@secondoftwo}%
\providecommand \href [0]{\begingroup \@sanitize@url \@href}%
\providecommand \@href[1]{\@@startlink{#1}\@@href}%
\providecommand \@@href[1]{\endgroup#1\@@endlink}%
\providecommand \@sanitize@url [0]{\catcode `\\12\catcode `\$12\catcode
  `\&12\catcode `\#12\catcode `\^12\catcode `\_12\catcode `\%12\relax}%
\providecommand \@@startlink[1]{}%
\providecommand \@@endlink[0]{}%
\providecommand \url  [0]{\begingroup\@sanitize@url \@url }%
\providecommand \@url [1]{\endgroup\@href {#1}{\urlprefix }}%
\providecommand \urlprefix  [0]{URL }%
\providecommand \Eprint [0]{\href }%
\providecommand \doibase [0]{http://dx.doi.org/}%
\providecommand \selectlanguage [0]{\@gobble}%
\providecommand \bibinfo  [0]{\@secondoftwo}%
\providecommand \bibfield  [0]{\@secondoftwo}%
\providecommand \translation [1]{[#1]}%
\providecommand \BibitemOpen [0]{}%
\providecommand \bibitemStop [0]{}%
\providecommand \bibitemNoStop [0]{.\EOS\space}%
\providecommand \EOS [0]{\spacefactor3000\relax}%
\providecommand \BibitemShut  [1]{\csname bibitem#1\endcsname}%
\let\auto@bib@innerbib\@empty
\bibitem [{\citenamefont {{Tarrataca}}\ \emph {et~al.}(2020)\citenamefont
  {{Tarrataca}}, \citenamefont {{Dias}}, \citenamefont {{Haddad}},\ and\
  \citenamefont {{Arruda}}}]{2020arXiv200406916T}%
  \BibitemOpen
  \bibfield  {author} {\bibinfo {author} {\bibfnamefont {L.}~\bibnamefont
  {{Tarrataca}}}, \bibinfo {author} {\bibfnamefont {C.~M.}\ \bibnamefont
  {{Dias}}}, \bibinfo {author} {\bibfnamefont {D.~B.}\ \bibnamefont
  {{Haddad}}}, \ and\ \bibinfo {author} {\bibfnamefont {E.~F.}\ \bibnamefont
  {{Arruda}}},\ }\href@noop {} {\bibfield  {journal} {\bibinfo  {journal}
  {arXiv e-prints}\ ,\ \bibinfo {eid} {arXiv:2004.06916}} (\bibinfo {year}
  {2020})},\ \Eprint {http://arxiv.org/abs/2004.06916} {arXiv:2004.06916
  [q-bio.PE]} \BibitemShut {NoStop}%
\bibitem [{\citenamefont {{Brethouwer}}\ \emph {et~al.}(2020)\citenamefont
  {{Brethouwer}}, \citenamefont {{van de Rijt}}, \citenamefont {{Lindelauf}},\
  and\ \citenamefont {{Fokkink}}}]{2020arXiv200406891B}%
  \BibitemOpen
  \bibfield  {author} {\bibinfo {author} {\bibfnamefont {J.-T.}\ \bibnamefont
  {{Brethouwer}}}, \bibinfo {author} {\bibfnamefont {A.}~\bibnamefont {{van de
  Rijt}}}, \bibinfo {author} {\bibfnamefont {R.}~\bibnamefont {{Lindelauf}}}, \
  and\ \bibinfo {author} {\bibfnamefont {R.}~\bibnamefont {{Fokkink}}},\
  }\href@noop {} {\bibfield  {journal} {\bibinfo  {journal} {arXiv e-prints}\
  ,\ \bibinfo {eid} {arXiv:2004.06891}} (\bibinfo {year} {2020})},\ \Eprint
  {http://arxiv.org/abs/2004.06891} {arXiv:2004.06891 [cs.SI]} \BibitemShut
  {NoStop}%
\bibitem [{\citenamefont {{Tsiotas}}\ and\ \citenamefont
  {{Magafas}}(2020)}]{2020arXiv200406536T}%
  \BibitemOpen
  \bibfield  {author} {\bibinfo {author} {\bibfnamefont {D.}~\bibnamefont
  {{Tsiotas}}}\ and\ \bibinfo {author} {\bibfnamefont {L.}~\bibnamefont
  {{Magafas}}},\ }\href@noop {} {\bibfield  {journal} {\bibinfo  {journal}
  {arXiv e-prints}\ ,\ \bibinfo {eid} {arXiv:2004.06536}} (\bibinfo {year}
  {2020})},\ \Eprint {http://arxiv.org/abs/2004.06536} {arXiv:2004.06536
  [physics.soc-ph]} \BibitemShut {NoStop}%
\bibitem [{\citenamefont {{Fowler}}\ \emph {et~al.}(2020)\citenamefont
  {{Fowler}}, \citenamefont {{Hill}}, \citenamefont {{Levin}},\ and\
  \citenamefont {{Obradovich}}}]{2020arXiv200406098F}%
  \BibitemOpen
  \bibfield  {author} {\bibinfo {author} {\bibfnamefont {J.~H.}\ \bibnamefont
  {{Fowler}}}, \bibinfo {author} {\bibfnamefont {S.~J.}\ \bibnamefont
  {{Hill}}}, \bibinfo {author} {\bibfnamefont {R.}~\bibnamefont {{Levin}}}, \
  and\ \bibinfo {author} {\bibfnamefont {N.}~\bibnamefont {{Obradovich}}},\
  }\href@noop {} {\bibfield  {journal} {\bibinfo  {journal} {arXiv e-prints}\
  ,\ \bibinfo {eid} {arXiv:2004.06098}} (\bibinfo {year} {2020})},\ \Eprint
  {http://arxiv.org/abs/2004.06098} {arXiv:2004.06098 [stat.AP]} \BibitemShut
  {NoStop}%
\bibitem [{\citenamefont {{Senapati}}\ \emph {et~al.}(2020)\citenamefont
  {{Senapati}}, \citenamefont {{Rana}}, \citenamefont {{Das}},\ and\
  \citenamefont {{Chattopadhyay}}}]{2020arXiv200404950S}%
  \BibitemOpen
  \bibfield  {author} {\bibinfo {author} {\bibfnamefont {A.}~\bibnamefont
  {{Senapati}}}, \bibinfo {author} {\bibfnamefont {S.}~\bibnamefont {{Rana}}},
  \bibinfo {author} {\bibfnamefont {T.}~\bibnamefont {{Das}}}, \ and\ \bibinfo
  {author} {\bibfnamefont {J.}~\bibnamefont {{Chattopadhyay}}},\ }\href@noop {}
  {\bibfield  {journal} {\bibinfo  {journal} {arXiv e-prints}\ ,\ \bibinfo
  {eid} {arXiv:2004.04950}} (\bibinfo {year} {2020})},\ \Eprint
  {http://arxiv.org/abs/2004.04950} {arXiv:2004.04950 [q-bio.PE]} \BibitemShut
  {NoStop}%
\bibitem [{\citenamefont {{Chikina}}\ and\ \citenamefont
  {{Pegden}}(2020)}]{2020arXiv200404144C}%
  \BibitemOpen
  \bibfield  {author} {\bibinfo {author} {\bibfnamefont {M.}~\bibnamefont
  {{Chikina}}}\ and\ \bibinfo {author} {\bibfnamefont {W.}~\bibnamefont
  {{Pegden}}},\ }\href@noop {} {\bibfield  {journal} {\bibinfo  {journal}
  {arXiv e-prints}\ ,\ \bibinfo {eid} {arXiv:2004.04144}} (\bibinfo {year}
  {2020})},\ \Eprint {http://arxiv.org/abs/2004.04144} {arXiv:2004.04144
  [q-bio.PE]} \BibitemShut {NoStop}%
\bibitem [{\citenamefont {{Amla}}\ and\ \citenamefont
  {{Amla}}(2020)}]{2020arXiv200403200A}%
  \BibitemOpen
  \bibfield  {author} {\bibinfo {author} {\bibfnamefont {K.}~\bibnamefont
  {{Amla}}}\ and\ \bibinfo {author} {\bibfnamefont {T.}~\bibnamefont
  {{Amla}}},\ }\href@noop {} {\bibfield  {journal} {\bibinfo  {journal} {arXiv
  e-prints}\ ,\ \bibinfo {eid} {arXiv:2004.03200}} (\bibinfo {year} {2020})},\
  \Eprint {http://arxiv.org/abs/2004.03200} {arXiv:2004.03200 [q-bio.PE]}
  \BibitemShut {NoStop}%
\bibitem [{\citenamefont {{Mohler}}\ \emph {et~al.}(2020)\citenamefont
  {{Mohler}}, \citenamefont {{Schoenberg}}, \citenamefont {{Short}},\ and\
  \citenamefont {{Sledge}}}]{2020arXiv200401714M}%
  \BibitemOpen
  \bibfield  {author} {\bibinfo {author} {\bibfnamefont {G.}~\bibnamefont
  {{Mohler}}}, \bibinfo {author} {\bibfnamefont {F.}~\bibnamefont
  {{Schoenberg}}}, \bibinfo {author} {\bibfnamefont {M.~B.}\ \bibnamefont
  {{Short}}}, \ and\ \bibinfo {author} {\bibfnamefont {D.}~\bibnamefont
  {{Sledge}}},\ }\href@noop {} {\bibfield  {journal} {\bibinfo  {journal}
  {arXiv e-prints}\ ,\ \bibinfo {eid} {arXiv:2004.01714}} (\bibinfo {year}
  {2020})},\ \Eprint {http://arxiv.org/abs/2004.01714} {arXiv:2004.01714
  [q-bio.PE]} \BibitemShut {NoStop}%
\bibitem [{\citenamefont {{Rizk-Allah}}\ and\ \citenamefont
  {{Hassanien}}(2020)}]{2020arXiv200405960R}%
  \BibitemOpen
  \bibfield  {author} {\bibinfo {author} {\bibfnamefont {R.~M.}\ \bibnamefont
  {{Rizk-Allah}}}\ and\ \bibinfo {author} {\bibfnamefont {A.~E.}\ \bibnamefont
  {{Hassanien}}},\ }\href@noop {} {\bibfield  {journal} {\bibinfo  {journal}
  {arXiv e-prints}\ ,\ \bibinfo {eid} {arXiv:2004.05960}} (\bibinfo {year}
  {2020})},\ \Eprint {http://arxiv.org/abs/2004.05960} {arXiv:2004.05960
  [cs.NE]} \BibitemShut {NoStop}%
\bibitem [{\citenamefont {{Bertozzi}}\ \emph {et~al.}(2020)\citenamefont
  {{Bertozzi}}, \citenamefont {{Franco}}, \citenamefont {{Mohler}},
  \citenamefont {{Short}},\ and\ \citenamefont
  {{Sledge}}}]{2020arXiv200404741B}%
  \BibitemOpen
  \bibfield  {author} {\bibinfo {author} {\bibfnamefont {A.~L.}\ \bibnamefont
  {{Bertozzi}}}, \bibinfo {author} {\bibfnamefont {E.}~\bibnamefont
  {{Franco}}}, \bibinfo {author} {\bibfnamefont {G.}~\bibnamefont {{Mohler}}},
  \bibinfo {author} {\bibfnamefont {M.~B.}\ \bibnamefont {{Short}}}, \ and\
  \bibinfo {author} {\bibfnamefont {D.}~\bibnamefont {{Sledge}}},\ }\href@noop
  {} {\bibfield  {journal} {\bibinfo  {journal} {arXiv e-prints}\ ,\ \bibinfo
  {eid} {arXiv:2004.04741}} (\bibinfo {year} {2020})},\ \Eprint
  {http://arxiv.org/abs/2004.04741} {arXiv:2004.04741 [q-bio.PE]} \BibitemShut
  {NoStop}%
\bibitem [{\citenamefont {{Villalobos-Arias}}(2020)}]{2020arXiv200402406V}%
  \BibitemOpen
  \bibfield  {author} {\bibinfo {author} {\bibfnamefont {M.}~\bibnamefont
  {{Villalobos-Arias}}},\ }\href@noop {} {\bibfield  {journal} {\bibinfo
  {journal} {arXiv e-prints}\ ,\ \bibinfo {eid} {arXiv:2004.02406}} (\bibinfo
  {year} {2020})},\ \Eprint {http://arxiv.org/abs/2004.02406} {arXiv:2004.02406
  [q-bio.PE]} \BibitemShut {NoStop}%
\bibitem [{\citenamefont {{Mbaye Ndiaye}}\ \emph {et~al.}(2020)\citenamefont
  {{Mbaye Ndiaye}}, \citenamefont {{Tendeng}},\ and\ \citenamefont
  {{Seck}}}]{2020arXiv200401574M}%
  \BibitemOpen
  \bibfield  {author} {\bibinfo {author} {\bibfnamefont {B.}~\bibnamefont
  {{Mbaye Ndiaye}}}, \bibinfo {author} {\bibfnamefont {L.}~\bibnamefont
  {{Tendeng}}}, \ and\ \bibinfo {author} {\bibfnamefont {D.}~\bibnamefont
  {{Seck}}},\ }\href@noop {} {\bibfield  {journal} {\bibinfo  {journal} {arXiv
  e-prints}\ ,\ \bibinfo {eid} {arXiv:2004.01574}} (\bibinfo {year} {2020})},\
  \Eprint {http://arxiv.org/abs/2004.01574} {arXiv:2004.01574 [q-bio.PE]}
  \BibitemShut {NoStop}%
\bibitem [{\citenamefont {{Dehning}}\ \emph {et~al.}(2020)\citenamefont
  {{Dehning}}, \citenamefont {{Zierenberg}}, \citenamefont {{Spitzner}},
  \citenamefont {{Wibral}}, \citenamefont {{Pinheiro Neto}}, \citenamefont
  {{Wilczek}},\ and\ \citenamefont {{Priesemann}}}]{2020arXiv200401105D}%
  \BibitemOpen
  \bibfield  {author} {\bibinfo {author} {\bibfnamefont {J.}~\bibnamefont
  {{Dehning}}}, \bibinfo {author} {\bibfnamefont {J.}~\bibnamefont
  {{Zierenberg}}}, \bibinfo {author} {\bibfnamefont {F.~P.}\ \bibnamefont
  {{Spitzner}}}, \bibinfo {author} {\bibfnamefont {M.}~\bibnamefont
  {{Wibral}}}, \bibinfo {author} {\bibfnamefont {J.}~\bibnamefont {{Pinheiro
  Neto}}}, \bibinfo {author} {\bibfnamefont {M.}~\bibnamefont {{Wilczek}}}, \
  and\ \bibinfo {author} {\bibfnamefont {V.}~\bibnamefont {{Priesemann}}},\
  }\href@noop {} {\bibfield  {journal} {\bibinfo  {journal} {arXiv e-prints}\
  ,\ \bibinfo {eid} {arXiv:2004.01105}} (\bibinfo {year} {2020})},\ \Eprint
  {http://arxiv.org/abs/2004.01105} {arXiv:2004.01105 [q-bio.PE]} \BibitemShut
  {NoStop}%
\bibitem [{\citenamefont {{Naresh Dhanwant}}\ and\ \citenamefont
  {{Ramanathan}}(2020)}]{2020arXiv200400696N}%
  \BibitemOpen
  \bibfield  {author} {\bibinfo {author} {\bibfnamefont {J.}~\bibnamefont
  {{Naresh Dhanwant}}}\ and\ \bibinfo {author} {\bibfnamefont {V.}~\bibnamefont
  {{Ramanathan}}},\ }\href@noop {} {\bibfield  {journal} {\bibinfo  {journal}
  {arXiv e-prints}\ ,\ \bibinfo {eid} {arXiv:2004.00696}} (\bibinfo {year}
  {2020})},\ \Eprint {http://arxiv.org/abs/2004.00696} {arXiv:2004.00696
  [q-bio.PE]} \BibitemShut {NoStop}%
\bibitem [{\citenamefont {{Alvarez}}(2020)}]{2020arXiv200310017A}%
  \BibitemOpen
  \bibfield  {author} {\bibinfo {author} {\bibfnamefont {L.}~\bibnamefont
  {{Alvarez}}},\ }\href@noop {} {\bibfield  {journal} {\bibinfo  {journal}
  {arXiv e-prints}\ ,\ \bibinfo {eid} {arXiv:2003.10017}} (\bibinfo {year}
  {2020})},\ \Eprint {http://arxiv.org/abs/2003.10017} {arXiv:2003.10017
  [q-bio.PE]} \BibitemShut {NoStop}%
\bibitem [{\citenamefont {{Hu}}\ \emph {et~al.}(2020)\citenamefont {{Hu}},
  \citenamefont {{Ge}}, \citenamefont {{Li}}, \citenamefont {{Boerwincle}},
  \citenamefont {{Jin}},\ and\ \citenamefont {{Xiong}}}]{2020arXiv200309800H}%
  \BibitemOpen
  \bibfield  {author} {\bibinfo {author} {\bibfnamefont {Z.}~\bibnamefont
  {{Hu}}}, \bibinfo {author} {\bibfnamefont {Q.}~\bibnamefont {{Ge}}}, \bibinfo
  {author} {\bibfnamefont {S.}~\bibnamefont {{Li}}}, \bibinfo {author}
  {\bibfnamefont {E.}~\bibnamefont {{Boerwincle}}}, \bibinfo {author}
  {\bibfnamefont {L.}~\bibnamefont {{Jin}}}, \ and\ \bibinfo {author}
  {\bibfnamefont {M.}~\bibnamefont {{Xiong}}},\ }\href@noop {} {\bibfield
  {journal} {\bibinfo  {journal} {arXiv e-prints}\ ,\ \bibinfo {eid}
  {arXiv:2003.09800}} (\bibinfo {year} {2020})},\ \Eprint
  {http://arxiv.org/abs/2003.09800} {arXiv:2003.09800 [q-bio.PE]} \BibitemShut
  {NoStop}%
\bibitem [{\citenamefont {{Fanelli}}\ and\ \citenamefont
  {{Piazza}}(2020)}]{2020arXiv200306031F}%
  \BibitemOpen
  \bibfield  {author} {\bibinfo {author} {\bibfnamefont {D.}~\bibnamefont
  {{Fanelli}}}\ and\ \bibinfo {author} {\bibfnamefont {F.}~\bibnamefont
  {{Piazza}}},\ }\href@noop {} {\bibfield  {journal} {\bibinfo  {journal}
  {arXiv e-prints}\ ,\ \bibinfo {eid} {arXiv:2003.06031}} (\bibinfo {year}
  {2020})},\ \Eprint {http://arxiv.org/abs/2003.06031} {arXiv:2003.06031
  [q-bio.PE]} \BibitemShut {NoStop}%
\bibitem [{\citenamefont {{Prince}}(2020)}]{prince2020}%
  \BibitemOpen
  \bibfield  {author} {\bibinfo {author} {\bibfnamefont {M.~e.}\ \bibnamefont
  {{Prince}}},\ }\href@noop {} {\bibfield  {journal} {\bibinfo  {journal}
  {arXiv e-prints}\ ,\ \bibinfo {eid} {arXiv:2020.04.09.20057091}} (\bibinfo
  {year} {2020})},\ \Eprint {http://arxiv.org/abs/2020.04.09.20057091}
  {medRXiv:2020.04.09.20057091} \BibitemShut {NoStop}%
\bibitem [{\citenamefont {Gumbel}(1935)}]{gumbel}%
  \BibitemOpen
  \bibfield  {author} {\bibinfo {author} {\bibfnamefont {E.}~\bibnamefont
  {Gumbel}},\ }\href@noop {} {\enquote {\bibinfo {title} {Les valeurs extremes
  des distributions statistiques},}\ } (\bibinfo {year} {1935})\BibitemShut
  {NoStop}%
\bibitem [{\citenamefont {{Bianconi}}\ \emph {et~al.}(2020)\citenamefont
  {{Bianconi}}, \citenamefont {{Marcelli}}, \citenamefont {{Campi}},\ and\
  \citenamefont {{Perali}}}]{2020arXiv200404604B}%
  \BibitemOpen
  \bibfield  {author} {\bibinfo {author} {\bibfnamefont {A.}~\bibnamefont
  {{Bianconi}}}, \bibinfo {author} {\bibfnamefont {A.}~\bibnamefont
  {{Marcelli}}}, \bibinfo {author} {\bibfnamefont {G.}~\bibnamefont {{Campi}}},
  \ and\ \bibinfo {author} {\bibfnamefont {A.}~\bibnamefont {{Perali}}},\
  }\href@noop {} {\bibfield  {journal} {\bibinfo  {journal} {arXiv e-prints}\
  ,\ \bibinfo {eid} {arXiv:2004.04604}} (\bibinfo {year} {2020})},\ \Eprint
  {http://arxiv.org/abs/2004.04604} {arXiv:2004.04604 [q-bio.PE]} \BibitemShut
  {NoStop}%
\bibitem [{\citenamefont {{\it Worldometer}}(2020)}]{worldo}%
  \BibitemOpen
  \bibfield  {author} {\bibinfo {author} {\bibnamefont {{\it Worldometer}}},\
  }\href@noop {} {\enquote {\bibinfo {title}
  {www.worldometers.info/coronavirus/},}\ } (\bibinfo {year}
  {2020})\BibitemShut {NoStop}%
\bibitem [{\citenamefont {{pcm-dpc data repository}}(2020)}]{pcmldo}%
  \BibitemOpen
  \bibfield  {author} {\bibinfo {author} {\bibnamefont {{pcm-dpc data
  repository}}},\ }\href@noop {} {\enquote {\bibinfo {title}
  {github.com/pcm-dpc/covid-19},}\ } (\bibinfo {year} {2020})\BibitemShut
  {NoStop}%
\bibitem [{\citenamefont {{CSS data repository}}(2020)}]{cssldo}%
  \BibitemOpen
  \bibfield  {author} {\bibinfo {author} {\bibnamefont {{CSS data
  repository}}},\ }\href@noop {} {\enquote {\bibinfo {title}
  {github.com/cssegisanddata/covid-19/tree/master/},}\ } (\bibinfo {year}
  {2020})\BibitemShut {NoStop}%
\bibitem [{\citenamefont {{Eco di Bergamo}}(2020)}]{ecoldo}%
  \BibitemOpen
  \bibfield  {author} {\bibinfo {author} {\bibnamefont {{Eco di Bergamo}}},\
  }\href@noop {} {\enquote {\bibinfo {title}
  {www.ecodibergamo.it/stories/bergamo-citta/quasi-mille-morti-nella-bergamascai-sindaci-ma-sono-molti-di-piu\_1346006\_11/},}\
  } (\bibinfo {year} {2020})\BibitemShut {NoStop}%
\bibitem [{\citenamefont {{Corriere della Sera}}(2020)}]{corldo}%
  \BibitemOpen
  \bibfield  {author} {\bibinfo {author} {\bibnamefont {{Corriere della
  Sera}}},\ }\href@noop {} {\enquote {\bibinfo {title}
  {www.corriere.it/politica/20\_marzo\_25/numero-vero-morti-covid-19-almeno-4-volte-quello-ufficiale-eebbe3ae-6eb8-11ea-925b-a0c3cdbe1130.shtml},}\
  } (\bibinfo {year} {2020})\BibitemShut {NoStop}%
\bibitem [{\citenamefont {{Pandemic Substack blog}}(2020)}]{panldo}%
  \BibitemOpen
  \bibfield  {author} {\bibinfo {author} {\bibnamefont {{Pandemic Substack
  blog}}},\ }\href@noop {} {\enquote {\bibinfo {title}
  {pandemic.substack.com/p/the-elephant-in-the-room-undercounting},}\ }
  (\bibinfo {year} {2020})\BibitemShut {NoStop}%
\bibitem [{\citenamefont {{Russel}}\ and\ \citenamefont
  {{al.}}(2020)}]{rusldo}%
  \BibitemOpen
  \bibfield  {author} {\bibinfo {author} {\bibfnamefont {T.~W.}\ \bibnamefont
  {{Russel}}}\ and\ \bibinfo {author} {\bibnamefont {{al.}}},\ }\href@noop {}
  {\enquote {\bibinfo {title}
  {\href{cmmid.github.io/topics/covid19/severity/diamond\_cruise\_cfr\_estimates.html}{cmmid.github.io/topics/covid19/severity/diamond\_cruise\_cfr\_estimates.html}},}\
  } (\bibinfo {year} {2020})\BibitemShut {NoStop}%
\bibitem [{\citenamefont {{Kermack}}(1927)}]{kermack27}%
  \BibitemOpen
  \bibfield  {author} {\bibinfo {author} {\bibfnamefont {A.~G.}\ \bibnamefont
  {{Kermack}}, \bibfnamefont {W.~O.;~{McKendrick}}},\ }\href@noop {} {\bibfield
   {journal} {\bibinfo  {journal} {Proceedings of the Royal Society A}\ }
  (\bibinfo {year} {1927})}\BibitemShut {NoStop}%
\bibitem [{\citenamefont {{Hethcote}}(2000)}]{hethcote00}%
  \BibitemOpen
  \bibfield  {author} {\bibinfo {author} {\bibfnamefont {H.}~\bibnamefont
  {{Hethcote}}},\ }\href@noop {} {\bibfield  {journal} {\bibinfo  {journal}
  {SIAM Review}\ }\textbf {\bibinfo {volume} {42 (4)}},\ \bibinfo {pages} {599}
  (\bibinfo {year} {2000})}\BibitemShut {NoStop}%
\bibitem [{har(2014)}]{harko14}%
  \BibitemOpen
  \href@noop {} {\bibfield  {journal} {\bibinfo  {journal} {Applied Mathematics
  and Computation}\ }\textbf {\bibinfo {volume} {236}},\ \bibinfo {pages} {184}
  (\bibinfo {year} {2014})},\ \Eprint {http://arxiv.org/abs/1403.2160}
  {arXiv:1403.2160} \BibitemShut {NoStop}%
\bibitem [{\citenamefont {{Bailey}}(1975)}]{bailey75}%
  \BibitemOpen
  \bibfield  {author} {\bibinfo {author} {\bibfnamefont {N.}~\bibnamefont
  {{Bailey}}},\ }\href@noop {} {\emph {\bibinfo {title} {The mathematical
  theory of infectious diseases and its applications (2nd ed.)}}}\ (\bibinfo
  {publisher} {London. Griffin},\ \bibinfo {year} {1975})\BibitemShut {NoStop}%
\bibitem [{\citenamefont {{Altizer}}(2006)}]{altizer06}%
  \BibitemOpen
  \bibfield  {author} {\bibinfo {author} {\bibfnamefont {C.}~\bibnamefont
  {{Altizer}}, \bibfnamefont {S.;~{Nunn}}},\ }\href@noop {} {\emph {\bibinfo
  {title} {Infectious diseases in primates: behavior, ecology and evolution.
  Oxford Series in Ecology and Evolution}}}\ (\bibinfo {year}
  {2006})\BibitemShut {NoStop}%
\bibitem [{\citenamefont {{Miller}}(2017)}]{miller17}%
  \BibitemOpen
  \bibfield  {author} {\bibinfo {author} {\bibfnamefont {J.}~\bibnamefont
  {{Miller}}},\ }\href {\doibase 10.1016/j.idm.2016.12.003} {\bibfield
  {journal} {\bibinfo  {journal} {Infectious Disease Modelling}\ }\textbf
  {\bibinfo {volume} {2 (1)}},\ \bibinfo {pages} {35} (\bibinfo {year}
  {2017})}\BibitemShut {NoStop}%
\bibitem [{\citenamefont {Levin}(1989)}]{hethcote89}%
  \BibitemOpen
  \bibinfo {editor} {\bibfnamefont {T.~G.~L.}\ \bibnamefont {Levin},
  \bibfnamefont {S.A.;~Hallam}},\ ed.,\ \href {\doibase
  doi:10.1007/978-3-642-61317-3_5} {\emph {\bibinfo {title} {Three Basic
  Epidemiological Models}}},\ \bibinfo {series} {Applied Mathematical Ecology.
  Biomathematics}\ No.~\bibinfo {number} {18}\ (\bibinfo  {publisher} {Berlin:
  Springer},\ \bibinfo {year} {1989})\BibitemShut {NoStop}%
\end{thebibliography}%


\providecommand{\noopsort}[1]{}\providecommand{\singleletter}[1]{#1}%
%


\end{document}